# Equilibration of complexes of DNA and H-NS proteins on charged surfaces: A coarse-grained model point of view.

Marc JOYEUX [(*)]

*Laboratoire Interdisciplinaire de Physique (CNRS UMR5588),*

*Université Joseph Fourier Grenoble 1, BP 87, 38402 St Martin d'Hères, France*

**Abstract** : The Histone-like Nucleoid Structuring protein (H-NS) is a nucleoid-associated protein, which is involved in both gene regulation and DNA compaction. Although it is a key player in genome organization by forming bridges between DNA duplexes, the precise structure of complexes of DNA and H-NS proteins is still not well understood. In particular, it is not clear whether the structure of DNA/H-NS complexes in the living cell is similar to that of complexes deposited on mica surfaces, which may be observed by AFM microscopy. A coarse-grained model, which helps getting more insight into this question, is described and analyzed in the present paper. This model is able of describing both the bridging of bacterial DNA by H-NS in the bulk and the deposition and equilibration of the complex on a charged surface. Simulations performed with the model reveal that a slight attraction between DNA and the charged surface is sufficient to let DNA/H-NS complexes reorganize from 3D coils to planar plasmids bridged by H-NS proteins similar to those observed by AFM microscopy. They furthermore highlight the antagonistic effects of the interactions between DNA and the surface. Indeed, increasing these interactions slows down the equilibration of naked plasmids on the surface but, on the other hand, enables a faster equilibration of DNA/H-NS complexes. Based on the distribution of the lifetimes of H-NS bridges and the time evolution of the number of *trans*-binding protein dimers during equilibration of the complexes on the surface, it is argued that the decrease of the equilibration time of the complex upon increase of the interaction strength between DNA and the surface is ascribable to the associated decrease of the probability to form new bridges between DNA and the proteins.

(*) email : Marc.Joyeux@ujf-grenoble.fr

# I – INTRODUCTION

In the cell, numerous proteins interact continuously with DNA to ensure biological functions such as transcription and replication control or genetic code maintenance. DNA and proteins are however very large size systems and their interactions can last milliseconds, minutes, or even more. From the computational point of view, it is therefore not feasible with nowadays computers to investigate these mechanisms with models that describe the system at the atomistic level. This has prompted the emergence of simplified models, which reduce the dimensionality of the problem by replacing groups of several (or many) atoms by single pseudo-atoms. These so-called "coarse-grained" models, where up to 15 DNA base pairs may be represented by a single site and proteins by a few ones or even by a solid, lack most of the details of atomistic models and are not appropriate for investigating specific interactions between precise DNA sequences and proteins. In contrast, they are sufficient to model certain non-specific (mostly electrostatic) DNA/protein interactions and allow for the numerical integration of quite long trajectories for rather large systems. This paper deals precisely with the development of a coarse-grained model aimed at describing the dynamics of complexes of bacterial DNA and Nucleoid Associated Proteins (NAPs) in the neighbourhood of (and on) a charged surface and the conclusions that can be drawn from the simulations.

While located in a clearly delineated structure called the *nucleoid*, the chromosomal DNA of bacteria is nevertheless not confined into an envelope-enclosed organelle. This is in sharp contrast with the DNA of eukaryotes, which is enclosed in the nucleus. Still, the genomic DNA of prokaryotes is highly compacted, like that of eukaryotes. For example, the volume of an *Escherichia coli* cell during non-exponential growth is approximately 1.5 $\mu m^3$ and the nucleoid occupies about 15% of this volume, that is, 0.2 to 0.3 $\mu m^3$ [1]. The Worm-Like-Chain model predicts that unconstrained circular polymer molecules with contour length $L$ and persistence length $\xi$ should form a random coil with gyration radius $R_g \approx \sqrt{L\xi/6}$ [2], that is, about 3.6 µm for the chromosomal DNA of *E. coli* with $L \approx 1.6$ mm and $\xi \approx 50$ nm. The volume of a homogeneous sphere with the same radius of gyration being $V = \frac{4}{3}\pi(\frac{5}{3})^{3/2} R_g^3$, that is, about 440 µm$^3$, the volume of the nucleoid is consequently more than 3 orders of magnitude smaller than that expected for the free coil. While it is generally admitted that four different mechanisms may contribute to the compaction of DNA in the nucleoid, namely (i) the association of Nucleoid Associated Proteins with DNA, (ii) DNA supercoiling resulting from the over- or under-winding of a DNA strand, (iii) macromolecular crowding owing to the cytoplasm, and (iv) neutralization of the charges carried by the DNA molecule by multivalent ions and certain DNA binding proteins, the importance of each mechanism and their possible synergies [3] for the formation of the prokaryotic nucleoid is still the matter of on-going debate.

This work focuses on the compaction of bacterial DNA by the Histone-like Nucleoid Structuring protein (H-NS). H-NS is a small protein (137 residues, 15.5 kDa), which is present at high levels (15000 to 20000 copies per cell) during exponential growth and the early stationary phase [4]. It plays a key role in global regulation by regulating several hundreds of genes [5,6] as well as in genome organization by forming bridges between DNA duplexes [7,8]. In solution and at not too high concentrations, H-NS is found predominantly as a very stable dimer [9]. The dimer has two C-terminal DNA-binding domains, which can bind either to two different DNA duplexes (*trans*-binding, bridging mode) or to two neighboring sites on the same DNA duplex (*cis*-binding, non-bridging mode) [7,8,10,11]. The role of H-NS in the compaction of bacterial DNA is still not well understood. Recent confocal microscopy experiments have shown that the volume of isolated nucleoids does not vary much when they are incubated with a high concentration of H-NS proteins, while further addition of a molecular crowding agent, like PEG molecules, leads to a more dramatic compaction of the chromosomal DNA, thus pointing to a possible synergy between these two mechanisms [12]. Moreover, optical trap-driven unzipping assays led Wiggins *et al.* [10] to the conclusion that the DNA loops observed by scanning force microscopy imaging [7] may result from the antagonistic effects of *trans*- and *cis*-binding. In order to gain more insight into the compaction dynamics, we recently proposed a coarse-grained model for the compaction of bacterial DNA by H-NS [13,14]. These simulations highlighted the fact that the compaction of DNA plasmids with a few thousands base pairs indeed results from the subtle equilibrium between several competing factors, like the deformation dynamics of the plasmid and the several binding modes of protein dimers to DNA [13]. They also pointed out that the conformations of DNA/H-NS complexes are rather different in the bulk and on a planar surface [13], thus suggesting that conformations observed on mica surfaces, which consist mainly of planar plasmid rings bridged by H-NS dimers [7], may differ significantly from those that prevail in living cells. This later point clearly deserves further investigation, since this hypothesis was arrived at on the basis of very preliminary calculations, where the plasmid was simply constrained to remain in the neighborhood of a plane. A realistic description of the interactions between DNA and the mica surface is needed to show that the conformations observed in scanning force microscopy experiments indeed result from the equilibration of globular DNA/H-NS complexes conformations on the mica surface.

The purpose of this paper is consequently threefold : (i) propose a complete coarse-grained model, which is able of describing both the bridging of bacterial DNA by H-NS in the bulk and the deposition and equilibration of the resulting complex on a charged surface, (ii) validate the model and its parametrization through the simulation of the deposition of naked plasmids on the charged surface, and (iii) analyse the results of simulations performed with complexes of DNA and proteins to derive conclusions regarding the equilibration of these complexes on charged surface. The results presented below reveal that a slight attraction between DNA and the charged surface is sufficient to let DNA/H-NS complexes reorganize from 3D coils to planar conformations similar to those observed by Dame and co-workers [7]. They furthermore highlight the antagonistic effects of the interaction between the plasmid and the surface. Indeed, it will be shown that increasing these interactions slows down the equilibration of naked plasmids on the surface but, on the other hand, enables a faster equilibration of DNA/H-NS complexes. Based on the distribution of the lifetimes of H-NS bridges and the time evolution of the number of *trans*-binding protein dimers during equilibration of the complexes on the surface, it will be concluded that the decrease of the equilibration time of the complex upon increase of the interaction strength between DNA and the surface is ascribable to the associated decrease of the probability to form new bridges between DNA and the proteins.

The remainder of this paper is consequently organized as follows. The model is first described in Sec. II. It is then validated in Sec. III by simulating the equilibration of naked plasmids on charged surfaces. These simulations are also used to parametrize the model adequately and get important results to compare later with. Results dealing with the equilibration of DNA/H-NS complexes on charged surfaces are next presented and discussed in Sec. IV, which is therefore the core of the paper. We finally conclude in Sec. V.

## II – DESCRIPTION OF THE MESOSCOPIC MODEL

Construction of the coarse-grained model was aimed at mimicking the formation, deposition and equilibration of DNA/H-NS complexes on mica surfaces at the same concentrations as in the experimental work of Dame *et al.* [7], that is, 85 ng of DNA with 450 ng of H-NS diluted in 10 µl H-NS BB. The model consists of one pUC19 plasmid and 224 H-NS dimers at a temperature $T = 298$ K ($1\,k_{\rm B}T \approx 2.48$ kJ/mol) enclosed between a half-sphere of radius $R_0 = 0.547$ μm and a charged plane (see Fig. 1). The fact that the model takes into account only one plasmid implies, of course, that it would fail to capture the eventual effect of crowding on compaction due to the interaction between several plasmids. This is, however, not a serious limitation for the purpose of discussing the work of Dame *et al.* [7], since in this work 1 μm$^3$ of solution contained on average only 3 plasmids and the deposition frequency was about one complex per μm$^2$ of the mica surface. Interactions between several plasmids were therefore not likely to play an important role.

As in previous work [13,15-18], DNA is modelled as a chain of beads, where each bead represents 15 DNA base pairs (bp). The 2686 bp pUC19 plasmid is thus modelled as a cyclic chain of 179 beads. Each H-NS dimer is in turn modelled as a chain of 3 beads, in order to keep approximately the same level of coarse-graining as for the plasmid molecule. Electric charges of $-12\,\bar{e}$, $4\,\bar{e}$, and $-8\,\bar{e}$, are placed at the centre of each DNA bead, H-NS terminal bead, and H-NS central bead, respectively ($\bar{e}$ denotes the absolute charge of the electron), and the beads describing DNA and the proteins interact through stretching, bending, electrostatic and excluded volume terms. Up to this point, the model is similar to the one described in our previous work [13]. Several terms were added to the model to take the interactions between molecular sites and the charged surface into account. More precisely, the total potential energy of the system, $V_{\rm total}$, is written in the form

$$V_{\rm total} = E_{\rm pot} + V_{\rm DNA/SURF} + V_{\rm PROT/SURF}, \tag{1}$$

where $E_{\rm pot}$ describes the potential energy of the system without the charged surface, $V_{\rm DNA/SURF}$ the interactions between the plasmid molecule and the charged surface, and $V_{\rm PROT/SURF}$ the interactions between H-NS molecules and the charged surface. The detailed expression of $E_{\rm pot}$ was provided in the Supporting Material of Ref. [13]. It is also summarized in the Supplemental Material of the present paper [19] for the sake of completeness. The two crucial features of $E_{pot}$ are the depth of the interaction well between DNA and protein beads and the bending rigidity of H-NS dimers. Experiments have shown that the change in enthalpy $\Delta H$ on forming a complex between DNA and a H-NS monomer is about 11 $k_B T$ at 298 K [20] and the value of the model parameter $\chi$ in Eq. S7 of [19] was fixed to $\chi = 0.15 k_B T$ in order to reproduce this value. Besides, it can be estimated that the bending rigidity $G$ of H-NS dimers is of the order of a few $k_B T$ [13,21] and it was shown that $G = 2 k_B T$ provides a fair agreement between simulated and observed conformations of complexes of DNA and proteins deposited on mica plates [13]. The value $G = 2 k_B T$ was consequently used for all simulations reported here.

Moreover, as was done in previous work [22-26], it was assumed that the interaction between molecular sites and the surface depends only on the distance between the site and the surface. This interaction can actually be decomposed into a sum of many interactions between the molecular site and the many charged sites located on the surface. The relative positions of the molecular and surface sites can be expressed in the cylindrical coordinate system, with the cylindrical axis perpendicular to the surface. By formally summing over all radial distances and angular positions, one creates a potential that depends only on the longitudinal coordinates, that is, the distance between the molecular site and the surface. While previous work [22-26] used the Weeks-Chandler-Anderson potential [27] to describe the interaction between DNA and the surface, it was instead modelled in the present work as the sum of attractive electrostatic terms and repulsive excluded volume terms, like the interaction between DNA and protein molecules [19]:

$$V_{\text{DNA/SURF}} = E_e^{(\text{DNA/SURF})} + E_{ev}^{(\text{DNA/SURF})}$$

$$E_e^{(\text{DNA/SURF})} = e_{\text{SURF}} e_{\text{DNA}} \sum_{k=1}^{n} H(z_k) \qquad (2)$$

$$E_{ev}^{(\text{DNA/SURF})} = \chi \left| \frac{e_{\text{SURF}}}{e_{\text{DNA}}} \right| \sum_{k=1}^{n} F(z_k),$$

where

$$H(z) = \frac{1}{4\pi\varepsilon z} \exp\left(-\frac{z}{r_D}\right), \tag{3}$$

and

if $z \leq 2^{3/2} a$ : $F(z) = 4\left(\left(\frac{2a}{z}\right)^4 - \left(\frac{2a}{z}\right)^2\right) + 1$

if $z > 2^{3/2} a$ : $F(z) = 0$ . (4)

In Eq. (2), $z_k$ denotes the elevation of the *k*th bead of the plasmid above the charged plane located at $z = 0$, while $e_{\text{SURF}}$ is an effective charge, which characterizes electrostatic interactions between the surface and the plasmid. As will be discussed shortly, the electrostatic properties of the surface were varied by assuming different values for $e_{\text{SURF}}$. The values of the other parameters are identical to those used in Ref. [13], that is, $a = 1.78$ nm, $r_D = 3.07$ nm, $\varepsilon = 80\varepsilon_0$, and $\chi = 0.15\, k_B T$.

It should be emphasized that the interaction between DNA and the mica surface is a rather complex mechanism, since it is very likely mediated by magnesium ions that bridge between the negative charges of the phosphate backbone of DNA and the negative mica sites. The Debye-Hückel potential in Eqs. (2)-(4) is therefore certainly a rather crude model of this interaction. Still, it has been argued that experimental conditions that lead to the complete equilibration of DNA strands on the charged surface, as is the case for the experiment of Dame and co-workers [7], are necessarily driven by nonspecific long-range forces, and not by short-range effects, which would rather lead to a large part of the molecule retaining its 3D conformation after deposition [28]. The exact expression for these nonspecific long-range forces is therefore probably not crucial, as long as (i) at a certain distance from the surface each segment of DNA experiences an attraction towards the surface and tends to minimize its free energy by moving closer to it, and (ii) the adsorbed molecules are not restricted in their lateral motion and can equilibrate completely. It will be shown in Sec. III that the interaction potential in Eqs. (2)-(4) fulfils these requirements for a certain range of values of $e_{SURF}$, but a similar result would certainly have been obtained with other expressions, like the Weeks-Chandler-Anderson potential [27]. The depth of the interaction potential therefore appears as the only key parameter of the model. It will be shown in Sec. III how simulations dealing with the deposition of naked plasmids on the charged surface help estimate the range of meaningful values for $e_{SURF}$ and validate the model.

At last, it was assumed that proteins cannot stick to the surface and that interactions between protein sites and the surface are purely repulsive. This is a rather strong assumption, because it is known that most proteins stick to many surfaces and that adsorption may be accompanied by structural rearrangement or overshooting kinetics and followed by free diffusion on the surface or protein aggregation [29]. Motivation for this choice was twofold: (i) keep the number of proteins at a reasonable level, and (ii) avoid introduction of an additional parameter in the model. Indeed, if proteins would have been allowed to adhere on the surface, then a much larger number thereof should have been included in the model in order to insure that the number of proteins capable of interacting with the plasmid remains reasonable. The already very lengthy simulations discussed below would consequently have become prohibitively long. Moreover, the strength of the interaction between H-NS proteins and the mica surface is not known and should have been introduced as a free parameter in the model, as was done above for $e_{SURF}$. Letting this parameter vary would again have increased computation time by an unacceptable factor. Most importantly, it will be argued at the end of Sec. IV that taking the attraction between the surface and proteins into account would most probably have the same effect as increasing the strength of the attraction between DNA and the surface. It would therefore bring nothing really new in the model and this argument validates *a posteriori* the choice of a repulsive potential.

More precisely, the interaction between H-NS proteins and the surface is taken in the form

$$V_{\text{PROT/SURF}} = \chi \sum_{j=1}^{P} \sum_{m=1}^{3} F'(Z_{jm}), \tag{5}$$

where $Z_{jm}$ denotes the elevation of bead $m$ of protein dimer $j$ above the charged plane and $F'(Z)$ is defined according to

$$\text{if } Z \leq Z_{\min} : F'(Z) = \chi \left( \left( \frac{Z_{\min}}{Z} \right)^4 - 2 \left( \frac{Z_{\min}}{Z} \right)^2 + 1 \right)$$

$$\text{if } Z > Z_{\min} : F'(Z) = 0, \tag{6}$$

where $Z_{\min} = 1.5$ nm.

The potential interaction energies between a single DNA bead and a single protein bead and the surface are shown in Fig. 2 as a function of the elevation of the center of the bead above the charged plane for an effective charge $e_{\text{SURF}} = -e_{\text{DNA}}/20$. Whatever the value of $e_{\text{SURF}}$, the interaction energy between a DNA bead and the surface is minimum at $z \approx 1.56$ nm, while the surface starts repelling protein beads for elevations smaller than $Z_{\min} = 1.5$ nm.

The final conformations of the complexes of DNA and H-NS proteins obtained at the end of the eight 20 ms simulations discussed in Ref. [13] (see for example the lowest plot in Fig. 1 of Ref. [13]) were translated above the charged surface, such that at time $t = 0$ the center of mass of the plasmid laid on the $x = y = 0$ axis and the elevation of the lowest bead was equal to $z = 7$ nm. The dynamics of the system was then investigated by integrating numerically the Langevin equations of motion with kinetic energy terms neglected. Practically, the updated position vector for each bead (whether DNA or protein), $\mathbf{r}^{(n+1)}$, was computed from the current position vector, $\mathbf{r}^{(n)}$, according to

$$\mathbf{r}^{(n+1)} = \mathbf{r}^{(n)} + \frac{\Delta t}{6\pi \eta a} \mathbf{F}^{(n)} + \sqrt{\frac{2 k_B T \Delta t}{6\pi \eta a}} \xi^{(n)}, \tag{7}$$

where $\Delta t = 20$ ps is the integration time step, $\mathbf{F}^{(n)}$ the collective vector of inter-particle forces arising from the potential energy $V_{\text{total}}$, $T = 298$ K the temperature of the system, $\xi^{(n)}$ a vector of random numbers extracted at each step $n$ from a Gaussian distribution of mean 0 and variance 1, and $\eta = 0.00089$ Pa s the viscosity of the buffer at 298 K. After each integration step, the position of the centre of the sphere was adjusted so as to coincide with the centre of mass of the DNA molecule.

In this model, the solvent is thus taken into account only through the Debye length $r_D$ (Eq. (3) and Eqs. (S2)-(S3) of the Supplemental Material [19]) and the damping coefficient $6\pi\eta a$ in Eq. (7). The Debye length describes the screening of electrostatic interactions between charges located on DNA, the proteins, and the charged surface, due to charges in the solvent. As a consequence, its value depends on the solvent and in particular its salinity. The damping coefficient $6\pi\eta a$ describes the effect of collisions between DNA and protein atoms and the solvent. The value of the hydrodynamic radius $a = 1.78$ nm was adjusted so as to reproduce the measured value of the diffusion coefficient of DNA [18]. The solvent is therefore taken into account only in a statistical way and more subtle effects, like for example hydration shells, are completely disregarded. Owing to the other approximations of the model, in particular the large coarse-graining, this is however probably of little consequence for the results.

Two different sets of simulations were performed. Proteins were discarded from the first set of simulations, in order to validate the model by testing its ability to describe the deposition of naked plasmids on a charged surface for a certain range of values of $e_{\text{SURF}}$. These simulations, which are discussed in Sec. III, were also used to point out the main characteristics of the equilibration of plasmids in the absence of proteins. Proteins were instead taken into account in the second set of simulations, which are discussed in Sec. IV, in order to (i) compare the principal features of the equilibration of the complexes on the surface with those of naked plasmids and (ii) provide an answer to the question raised in this paper concerning the hypothesis that the planar conformations observed with an AFM microscope might actually result from the equilibration of 3D globular complexes on the mica surface. Eight trajectories were run for each set of simulations and for different values of the effective surface charge $e_{\text{SURF}}$ ranging from $-e_{\text{DNA}}/25$ to $-e_{\text{DNA}}/5$, in order to check the influence of the interaction strength with the surface on the deposition and equilibration kinetics.

Analysis of the simulations consisted in tracking every 10 µs the number of bound H-NS dimers. It was considered that an H-NS dimer is bound to DNA if it has at least one bead which centre is within a distance of 2 nm from the centre of a DNA bead. It was shown in Ref. [13] that there is actually not much freedom in the choice of the distance threshold Indeed, for thresholds smaller than 2.0 nm, many bound proteins are missed because of the steep repulsive wall that surrounds DNA beads while, on the other hand, a certain number of dangling dimers are incorrectly reassigned as *cis*-binding dimers when the threshold is increased from 2.0 to 2.5 nm. The nature of the bound H-NS dimers was also tracked. A dimer was considered to be bound in *cis* if the two terminal beads were interacting with DNA beads separated by at most two DNA beads. If the H-NS dimer was instead interacting with two DNA beads separated by (strictly) more than two DNA beads, the H-NS/DNA interaction was counted as a *trans* bond. Finally, the dimer was considered to be dangling, if only one bead of the dimer was interacting with the DNA. The delineation between these three different cases is schematized in Fig. 3.

The radius of gyration of the plasmid molecule, $R_g$, as well as its aplanarity coefficient, $a$, were also computed every 10 µs. $R_g$ and $a$ are defined according to

$$R_g = (\lambda_1 + \lambda_2 + \lambda_3)^{1/2}$$
$$a = \frac{9\lambda_1\lambda_2\lambda_3}{(\lambda_1 + \lambda_2 + \lambda_3)(\lambda_1\lambda_2 + \lambda_2\lambda_3 + \lambda_3\lambda_1)} \quad (8)$$

in terms of the eigenvalues $\lambda_j$ ($j=1,2,3$) of the gyration tensor of the plasmid. $a$ varies between 0 and 1 and is equal to 0 for planar conformations.

## III – EQUILIBRATION OF NAKED PLASMIDS ON THE CHARGED SURFACE

Let us first examine the deposition and equilibration of naked plasmids on the charged surface. These simulations are used to validate the model and determine the range of meaningful values for the key parameter $e_{\text{SURF}}$, which governs the depth of the interaction between DNA and the surface (Eqs. (2)-(4)). They also provide a certain number of important results to which simulations with DNA/H-HS complexes will be compared in Sec. IV.

As described in Sec. II, at the beginning of each simulation the plasmid is positioned 7 nm above the surface and left free to evolve under the action of internal and external forces and thermal noise. Since in the model the interaction energy between DNA beads and the surface is minimum at a height $z \approx 1.56$ nm above the surface (see Fig. 2), the plasmid is attracted by the surface and spreads at the corresponding height. This is clearly seen in the left column of Fig. 4, where the top vignette shows the conformation of a naked plasmid at time $t=0$ and the middle one its conformation 0.3 ms later, just after it has completed adsorption on the surface. At this time, the plasmid is however not fully equilibrated, since duplexes overlap at two different places. It takes 5 additional milliseconds before these overlaps disappear as a consequence of random motions driven by thermal noise (see the bottom vignette in the left column of Fig. 4). After equilibration, the plasmid merely diffuses on the surface with a diffusion coefficient that does not depend on the strength of the interaction between DNA and the surface, that is, the effective charge $e_{\text{SURF}}$.

The simulation shown in the left column of Fig. 4 was performed with $e_{\text{SURF}} = -0.08$ $e_{\text{DNA}}$. Similar results were obtained for all values of $-e_{\text{SURF}}/e_{\text{DNA}}$ ranging from 0.04 to 0.10.

For $-e_{SURF}/e_{DNA} = 0.04$, the average binding energy between DNA and the surface is of the order of only $0.6\,k_BT$ per bead and the plasmid sometimes forms arches that protrude from the surface. For values of $-e_{SURF}/e_{DNA}$ smaller than 0.04, the plasmid does not adhere to the surface, while the plasmid sticks firmly to it for larger ones. Moreover, it was found that in most simulations the plasmid does not equilibrate completely on the surface for values of $-e_{SURF}/e_{DNA}$ equal to or larger than 0.13. More precisely, the 8-like geometry appears to be a stable plasmid conformation for such values of $-e_{SURF}/e_{DNA}$, while in a few cases equilibration was achieved through the simultaneous cancellation of the two overlaps of 8– like conformations. At this point, it is worth mentioning that Dame *et al.* [7] deposited the complexes of DNA and proteins on freshly cleaved mica surfaces without prior treatment, that is, in conditions where the density of surface charges is moderate and DNA molecules are known to equilibrate on the surface as in an ideal solution [30].

These simulations therefore show that the model described in Sec. II is indeed able to describe the adsorption and equilibration of plasmids on charged surfaces. They furthermore indicate that values of $-e_{SURF}/e_{DNA}$ appropriate to model the experiments in Ref. [7] range from 0.04 to about 0.10.

Simulations performed with naked plasmids also provide a set of important results to compare later with. To this end, one may notice that the adsorption, equilibration, and diffusion steps can be clearly monitored by plotting the time evolution of the radius of gyration $R_g$ and aplanarity coefficient $a$. Such plots are shown in Fig. 5 for the simulation already illustrated in the left column of Fig. 4. During adsorption, the plasmid transforms from a 3D coil to a rather flat, nearly 2D object. The aplanarity coefficient $a$ consequently decreases substantially during this phase (see the bottom plot of Fig. 5). Moreover, during equilibration, the shape of the plasmid transforms from an 8-like (or more complex) geometry to a ring-like geometry with no overlaps. The radius of gyration $R_g$ consequently increases markedly during this phase (see the top plot of Fig. 5). At last, fluctuations of $a$ are exceedingly small during the free diffusion phase, because nothing prevents the plasmid from remaining totally flat (see the bottom plot of Fig. 5).

The time evolution of $R_g$ and $a$ is even clearer when averaged over several trajectories. This can be checked in Fig. 6, where the evolution of $R_g$ and $a$ has been averaged over 8 simulations with different initial conditions but the same value $e_{SURF} = -0.08 e_{DNA}$. The characteristic times for adsorption, $\tau_a$, and equilibration, $\tau_R$, can be derived from these curves by fitting them with the expressions $a = a_0 \exp(-t/\tau_a)$ and $R_g = R_0 + \Delta R(1 - \exp(-t/\tau_R))$, respectively. The evolution of $\tau_a$ and $\tau_R$ as a function of $-e_{SURF}/e_{DNA}$ is further plotted in Fig. 7. It is seen in this figure that, except for the lowest value $e_{SURF} = -0.04 e_{DNA}$, for which the plasmid sometimes does not adhere completely to the surface, $\tau_R$ is always substantially larger than $\tau_a$. Moreover, $\tau_a$ is a decreasing function of $-e_{SURF}/e_{DNA}$. This is not surprising, because the force that attracts DNA towards the surface is larger for larger values of $-e_{SURF}/e_{DNA}$ and it consequently takes less time for the plasmid to adsorb on it. In contrast, the characteristic time for equilibration, $\tau_R$, is an increasing function of $-e_{SURF}/e_{DNA}$. This is due to the fact that equilibration requires that the plasmid transiently detaches from the surface to remove duplex overlaps and this motion becomes more difficult with increasing attraction between DNA and the surface. Last but not least, it should be noted that the average value of the radius of gyration of plasmids equilibrated on the surface does not depend on the value of $e_{SURF}$, ranging between 115 and 120 nm for all values of $-e_{SURF}/e_{DNA}$ comprised between 0.04 and 0.10.

## IV – EQUILIBRATION OF DNA/H-NS COMPLEXES ON THE CHARGED SURFACE

Let us now examine the deposition and equilibration of complexes of plasmid and H-NS proteins on the charged surface. This section is the core of the paper, since this work was motivated by the need to check the hypothesis that the planar conformations observed with an AFM microscope might actually result from the equilibration of 3D globular complexes on the mica surface.

As illustrated in the right column of Fig. 4, simulations performed with DNA/H-NS complexes display the same steps as simulations performed with naked plasmids. The complex first adsorbs on the surface, with duplexes overlapping at one or more places. It then takes some time before these overlaps are removed by random motions driven by thermal noise. At last, the equilibrated complex just diffuses freely on the surface. The proteins bound to the plasmid are however responsible for the fact that the complex adheres firmly to the surface only for values of $-e_{\text{SURF}}/e_{\text{DNA}}$ equal to or larger than 0.06 (against 0.05 for the naked plasmid). For $-e_{\text{SURF}}/e_{\text{DNA}} = 0.05$, the complex often forms arches that protrude from the surface, as is also the case for the naked plasmid and $-e_{\text{SURF}}/e_{\text{DNA}} = 0.04$.

These simulations therefore confirm the hypothesis that the planar conformations of DNA/H-NS complexes observed with an AFM microscope [7] probably result from the equilibration of 3D globular complexes on the mica surface [13]. Additional information on the equilibration process can further be gained by comparing more thoroughly the results with those obtained for naked plasmids.

As for naked plasmids, the three successive steps indeed leave clear fingerprints in the evolution of $R_g$ and $a$, with the aplanarity coefficient $a$ decreasing substantially during the adsorption phase and remaining very close to zero during the free diffusion phase, and the radius of gyration $R_g$ increasing markedly during the equilibration phase (see Fig. 8). Because of the several bridges between different duplexes formed by *trans*-binding H-NS dimers, the average radius of gyration of equilibrated DNA/H-NS complexes is however smaller than that of equilibrated naked plasmids, being close to 100 nm against about 115 nm for naked plasmids.

As for naked plasmids, the characteristic times for adsorption, $\tau_a$, and equilibration, $\tau_R$, were also obtained from the time evolution of $a$ and $R_g$ averaged over 8 simulations run for 40 ms (see Fig. 9). Note, however, that for $e_{SURF} = -0.06\, e_{DNA}$, the estimated value $\tau_R = 31$ ms must be considered as a rough estimate since the statistics is not sufficient to warrant good accuracy. Moreover, for $e_{SURF} = -0.05\, e_{DNA}$, $\tau_R$ is probably of the order of 100 ms or more but it can hardly be estimated from 40 ms simulations.

The evolution of $\tau_a$ and $\tau_R$ as a function of $-e_{SURF}/e_{DNA}$ is plotted in Fig. 10. Comparison of Figs. 7 and 10 indicates that $\tau_a$ and $\tau_R$ are always larger for DNA/H-NS complexes than for naked plasmids. The reason is, of course, that H-NS bridges slow down both the flattening of the complex on the surface and its subsequent equilibration, because these two mechanisms require breaking or reorganization of a large part of the bridges connecting different duplexes.

For values of $-e_{SURF}/e_{DNA}$ ranging from 0.10 down to 0.06, $\tau_a$ is about 5 times larger for DNA/H-NS complexes than for naked plasmids and this ratio increases up to more than 20 for the lowest value $-e_{SURF}/e_{DNA} = 0.05$ (10 ms against 0.45 ms). $\tau_a$ however remains a decreasing function of $-e_{SURF}/e_{DNA}$ for the adsorption of DNA/H-NS complexes, as is also the case for naked plasmids. Once again, this is not surprising, because the force that attracts DNA/H-NS complexes towards the surface is larger for larger values of $-e_{SURF}/e_{DNA}$ and it consequently takes less time for complexes to adsorb on it.

The ratios between the values of $\tau_R$ for DNA/H-NS complexes and naked plasmids are of the same order of magnitude as the ratios for $\tau_a$, increasing from about 2.5 for $-e_{SURF}/e_{DNA}=0.10$ (6.6 ms against 2.7 ms) to about 20 for $-e_{SURF}/e_{DNA}=0.06$ (31 ms against 1.8 ms). Quite interestingly, it can however be noticed in the bottom plot of Fig. 10 that $\tau_R$ is a decreasing function of $-e_{SURF}/e_{DNA}$ for DNA/H-NS complexes, while it is an increasing function for naked plasmids (see Fig. 7). The attraction force between the surface and the DNA therefore plays two antagonistic roles. On one side, it slows down the gliding of different duplexes with respect to one another, therefore making equilibration of naked plasmids increasingly slower and more difficult with increasing values of $e_{SURF}$ (Fig. 7), but on the other side it makes equilibration of DNA/H-NS complexes faster with increasing values of $e_{SURF}$ (bottom plot of Fig. 10). The remainder of this section is devoted to understanding why an increase of the interaction strength between DNA and the surface leads to a decrease of the equilibration time of the complex.

The first reason one might think of is the possibility that the attraction between DNA and the surface might help breaking the H-NS bridges that link different DNA duplexes and slow down the flattening and equilibration of the complex on the surface. This hypothesis can be checked by plotting the probability distribution of the lifetimes of H-NS bridges for different values of $-e_{SURF}/e_{DNA}$. Such a plot is shown in Fig. 11 for $-e_{SURF}/e_{DNA}=0.06$ and $-e_{SURF}/e_{DNA}=0.10$. This figure indicates clearly that the strength of the interaction between DNA and the surface has no significant effect on the lifetime of H-NS bridges.

To gain more detailed insight into the equilibration dynamics of DNA/H-NS complexes on charged surfaces and understand the antagonistic roles played by the interaction between DNA and the surface, it proved actually very helpful to plot the time evolution of the number of proteins bound to DNA in *trans*-, *cis*-, and dangling conformations. Such plots, obtained by averaging the number of respective conformations over 8 simulations with different initial conditions, are shown in Fig. 12 for the case $e_{\text{SURF}} = -0.08 e_{\text{DNA}}$ (similar plots were obtained for all values of $-e_{\text{SURF}}/e_{\text{DNA}}$ ranging from 0.05 to 0.10). Also shown on the same figure are adjusted functions that reproduce precisely the time evolution of the number of bound proteins. The basis sets for the fits are $\{1, \exp(-t/\tau_a), \exp(-t/\tau_R)\}$ for *trans*- and *cis*- conformations, and $\{1, t, \exp(-t/\tau_a), \exp(-t/\tau_R)\}$ for dangling proteins. The simplest time evolution is that of the number of *cis*-binding proteins. As can be checked in the middle plot of Fig. 12, it merely consists of a sharp decrease with time scale $\tau_a$ followed by a plateau at larger times. In contrast, it is seen in the bottom plot of Fig. 12 that the number of dangling proteins initially decreases with time scale $\tau_R$ before increasing again linearly with time. At last, the number of bridges between different duplexes (*i.e.* of *trans*-binding proteins) increases with time scale $\tau_a$ before decreasing with time scale $\tau_R$ (see the top plot of Fig. 12). These evolutions can be interpreted as follows. The adsorption of the complex on the surface is responsible for the deletion of a certain number of *cis* conformations, while the subsequent equilibration of the complex has no impact on *cis*-binding proteins. In contrast, adsorption on the surface has little impact on the highly mobile dangling proteins, while a rather large number of protein dimers detach from the plasmid during the equilibration phase, probably due to thermally driven stochastic motions of both DNA and the proteins. Later, dimers that collide with the plasmid while diffusing freely in the confining sphere replace, at a linear rate, the previously detached dangling protein dimers. However, most interesting to us is the dynamics of *trans*-binding proteins. As the complex crashes on the surface, duplexes are brought to shorter distances from one another, thus transiently favouring the formation of new bridges. In contrast, the number of *trans*-binding H-NS proteins then decreases during equilibration. As for dangling proteins, this is due to thermally driven stochastic motions of both DNA and the proteins, which break some of the bonds that formed between DNA and the proteins. The fact that this decrease occurs with rate $1/\tau_R$, that is, the same rate as the increase of $R_g$, indicates that these two mechanisms are intimately related and that reducing the number of H-NS bridges is indeed crucial for faster equilibration. The question "why does an increase of $-e_{SURF}/e_{DNA}$ lead to faster equilibration ?" is therefore equivalent to "why does an increase of $-e_{SURF}/e_{DNA}$ lead to a faster decrease of the number of H-NS bridges ?". As shown above, the lifetime of the bridges does not depend on the value of $-e_{SURF}/e_{DNA}$. The fact that the number of bridges decreases more rapidly with increasing values of $-e_{SURF}/e_{DNA}$ must therefore be attributed to the fact it becomes more and more difficult to form new bridges with increasing values of $-e_{SURF}/e_{DNA}$. This, in turn, does make sense, because DNA and the proteins interact through long range electrostatic interactions. If a DNA and a protein molecule come sufficiently close to feel their mutual attraction and are completely free to move, they will tend to come still closer and closer, in order to maximize the interaction energy. Instead, if their motion is hindered, then thermal noise may again separate them after a while. Reduction of the degrees of freedom of the molecules therefore leads to a decrease of the probability of forming a bond or a bridge. Now, increasing $-e_{SURF}/e_{DNA}$ amounts precisely to hindering more and more the motion of the DNA perpendicular to the surface, thereby reducing the probability to form a bond or a bridge with a neighbouring protein molecule. In conclusion, an increase of the interaction strength between DNA and the surface leads to a decrease of the equilibration time of the complex because of the associated decrease of the probability to form new bridges between DNA and the proteins.

Before closing this section, it may be worth coming back briefly to one of the assumptions of the model, namely that the interaction between proteins and the surface is uniquely repulsive (see Eqs. (5)-(6)). Since the protein concentrations used in the work of Dame and coworkers are far from being sufficient for surface coverage and proteins additionally diffuse rapidly on surfaces [31-33] compared to the much heavier DNA/H-NS complexes, eventual H-NS proteins adhering to the surface would not be able to hinder seriously the adsorption of the complex on the surface. Moreover, allowing proteins that interact with the adsorbed complex to stick to the surface would most probably essentially have the same effect as increasing the strength of the interaction between DNA and the surface. This would indeed reduce the mobility of these proteins perpendicular to the surface, thereby reducing the probability to form bonds and bridges between DNA duplexes and consequently accelerating somewhat the equilibration dynamics.

**V – CONCLUSION**

This paper proposes a model, which is able of describing both the bridging of bacterial DNA by H-NS in the bulk and the deposition and equilibration of the complex on a charged surface. Results of the simulations performed with this model reveal that even a rather weak attraction is sufficient to drive a drastic reorganization of the DNA/H-NS complex from a 3D coil to a flat ring with protein bridges similar to those observed by Dame and co-workers in AFM experiments [7,8,10]. This confirms the hypothesis that conformations observed on mica surfaces probably differ significantly from those that prevail in the living cell [13]. Moreover, comparison of the simulations performed for naked plasmids and DNA/H-NS complexes highlight the fact that the interactions between the plasmid and the surface actually have two antagonistic effects. On one hand, these interactions slow down the relative gliding of overlapping duplexes and consequently the equilibration of naked plasmids on the surface while, on the other hand, they reduce the probability to form new H-NS bridges, thereby facilitating the equilibration of DNA/H-NS complexes on the surface.

At this point, it is worth noting that the models proposed in Ref. [13] and the present paper predict only a mild compaction of bacterial DNA by H-NS proteins. The average radius of gyration of naked pUC19 plasmids is indeed close to 90 nm and that of the model of DNA/H-NS complexes discussed here close to 60 nm (prior to deposition on the charged surface). This conclusion agrees with both the previous simulations of de Vries performed with a different model [34] and the experimental observation that the 3 orders of magnitude compaction of bacterial DNA in the nucleoid cannot be due to the mere action of Nucleoid Associated Proteins (NAP) [12]. There are consequently several lines along which the model described in Ref. [13] and the present paper could be extended to investigate the compaction of bacterial DNA in more detail. For example, it is known that DNA supercoiling also contributes to condensation. The circular bacterial chromosome is indeed twisted by ATP-consuming enzymes, leading to branched plectonemic supercoils [35] that are more compact than coiled DNA. While it is probable that supercoiling cannot compact DNA by more than one order of magnitude [36], it would nevertheless be interesting to introduce torsion and twisting in our model according to the procedure described in Refs. [37,38] in order to check the level of DNA compaction predicted by a model for DNA with torsional degrees of freedom and the existence of possible synergies between NAP and supercoiling. Another possible extension of this work consists in performing simulations close to *in vivo* conditions instead of the experimental conditions of Dame *et al.* [7]. Bacterial DNA may indeed be several million base pairs long, against 2686 bp for the pUC19 plasmid, and each cell contains approximately one H-NS dimer per 200 DNA base pairs, while the experiments of Dame *et al.* were performed with one H-NS dimer per 12 DNA base pairs. It would be interesting to check whether *in vivo* conditions lead to the same compaction characteristics as the experiments of Dame *et al.* or to substantially different ones. At last, one could also introduce crowding agents in the model, in order to check the possible existence of synergies between NAP and crowding agents [12] or crowding agents and supercoiling. The two last suggestions are of course rather expensive from the computational point of view.

# REFERENCES

[1] C.L. Woldringh and N. Nanninga, in *Molecular cytology of Escherichia coli*, edited by N. Nanninga, (Academic Press, London, 1985), pp. 161-197.

[2] D.R. Latulippe and A.L. Zydney, Biotechnol. Bioeng. 107, 134 (2010).

[3] R. de Vries, Biochimie 92, 1715 (2010).

[4] T.A. Azam, A. Iwata, A. Nishimura, S. Ueda, and A. Ishihama, J. Bacteriol. 181, 6361 (1999).

[5] T. Atlung and H. Ingmer, Mol. Microbiol. 24, 7 (1997).

[6] F. Hommais, E. Krin, C. Laurent-Winter, O. Soutourina, A. Malpertuy, J.-P. Le Caer, A. Danchin, and P. Bertin, Mol. Microbiol. 40, 20 (2001).

[7] R.T. Dame, C. Wyman, and N. Goosen, Nucleic Acids Res. 28, 3504 (2000).

[8] R.T. Dame, M.C. Noom, and G.J.L. Wuite, Nature 444, 387 (2006).

[9] M. Falconi, M.T. Gualtieri, A. La Teana, M.A. Losso, and C.L. Pon, Mol. Microbiol. 2, 323 (1988).

[10] P.A. Wiggins, R.T. Dame, M.C. Noom, and G.J.L. Wuite, Biophys. J. 97, 1997 (2009).

[11] Y.J. Liu, H. Chen, L.J. Kenney, and J. Yan, Gene Dev. 24, 339 (2010).

[12] K. Wintraecken, *Interplay between the bacterial nucleoid protein H-NS and macromolecular crowding in compacting DNA* (PhD thesis, Wageningen University, Netherlands, 2012).

[13] M. Joyeux and J. Vreede, Biophys. J. 104, 1615 (2013).

[14] G.S. Freeman and J.J. de Pablo, Biophys. J. 104, 1397 (2013).

[15] A.-M. Florescu and M. Joyeux, J. Chem. Phys. 130, 015103 (2009).

[16] A.-M. Florescu and M. Joyeux, J. Chem. Phys. 131, 105102 (2009).

[17] A.-M. Florescu and M. Joyeux, J. Phys. Chem. A 114, 9662 (2010).

[18] H. Jian, A. Vologodskii, and T. Schlick, J. Comp. Phys. 136, 168 (1997).

[19] See supplemental material at [URL will be inserted by AIP] for a detailed description of the model without the charged surface and the expression of $E_{\text{pot}}$.

[20] S. Ono, M.D. Goldberg, T. Olsson, D. Esposito, J.C.D. Hinton, and J.E. Ladbury, Biochem. J. 391, 203 (2005).

[21] S. Sivaramakrishnan, J. Sung, M. Ali, S. Doniach, H. Flyvbjerg, and J.A. Spudich, Biophys. J. 97, 2993 (2009).

[22] T.J. Schmitt, J. B. Rogers, and T.A. Knotts IV, J. Chem. Phys. 138, 035102 (2013).

[23] T.J. Schmitt and T.A. Knotts IV, J. Chem. Phys. 134, 205105 (2011).

[24] T.A. Knotts IV, N. Rathore, and J.J. de Pablo, Proteins 61, 385 (2005).

[25] S. Wei and T.A. Knotts IV, J. Chem. Phys. 133, 115102 (2010).

[26] S. Wei and T.A. Knotts IV, J. Chem. Phys. 134, 185101 (2011).

[27] J.D. Weeks, D. Chandler, and H.C. Andersen, J. Chem. Phys. 544, 5237 (1971).

[28] M.L. Sushko, A.L. Shluger, and C. Rivetti, Langmuir 22, 7678 (2006).

[29] M. Rabe, D. Verdes, and S. Seeger, Adv. Colloid Interface Sci. 162, 87 (2011)

[30] C. Rivetti, M. Guthold, and C. Bustamante, J. Mol. Biol. 264, 919 (1996).

[31] D.T. Kim, H.W. Blanch, and C.J. Radke, Langmuir 18, 5841 (2002)

[32] K. Kubiak-Ossowska and P.A. Mulheran, Langmuir 28, 15577 (2012)

[33] T. Wei, M.A. Carignano, and I. Szleifer, J. Phys. Chem. B 116, 10189 (2012)

[34] R. de Vries, J. Chem. Phys. 135, 125104 (2011)

[35] C.R. Calladine, H.R. Drew, B.F. Luisi, and A.A. Travers, *Understanding DNA* (Elsevier Academic Press, Amsterdam, 2004), pp. 116-138.

[36] S. Cunha, C.L. Woldringh, and T. Odijk, J. Struct. Biol. 136, 53 (2001).

[37] G. Chirico and J. Langowski, Biopolymers 34, 415 (1994).

[38] K. Klenin, H. Merlitz, and J. Langowski, Biophys. J. 74, 780 (1998).

**FIGURE LEGENDS**

**Figure 1** : This figure shows the initial configuration for one of the simulations discussed in the text, with one plasmid (represented by a cyclic chain of 179 bead), 224 H-NS protein dimers (each dimer is represented by a linear chain of 3 beads), a charged plan, and a confining half-sphere of radius 0.547 µm. At time $t=0$ the elevation of the lowest DNA bead above the surface is equal to 7 nm.

**Figure 2 :** Plot of the potential interaction energy between a DNA bead and the surface (solid line) and a protein bead and the surface (dashed line) as a function of the elevation $z$ of the center of the bead above the charged plane, for an effective charge $e_{\text{SURF}} = -e_{\text{DNA}}/20$. Distances are expressed in nm and energies in units of $k_{\text{B}}T$.

**Figure 3 :** Diagram showing the three different ways H-NS proteins bind to DNA and their delineations. DNA and proteins are represented by dark and white disks, respectively.

**Figure 4** : This figure illustrates the dynamics of a naked plasmid (left column) and a complex of plasmid and H-NS proteins (right column) in the neighbourhood of the charged surface (represented as a disk). Top vignettes show the initial conformations of the molecules at an elevation of 7 nm above the plane. Middle vignettes show their conformations just after completing adsorption on the surface. Bottom vignettes show their conformations as they finally fully equilibrate. These simulations were performed with $e_{\text{SURF}} = -0.08\, e_{\text{DNA}}$.

**Figure 5 :** Plot of the time evolution of the radius of gyration $R_g$ (top plot) and the aplanarity coefficient $a$ (bottom plot) for the simulation with a naked plasmid and $e_{\text{SURF}} = -0.08\, e_{\text{DNA}}$

also shown in the left column of Fig. 4. The two dot-dashed vertical lines labeled $t_{ad}$ and $t_{eq}$ indicate the times where the plasmid has completed adsorption (see the middle vignette in the left column of Fig. 4) and equilibration (see the bottom vignette in the left column of Fig. 4), respectively.

**Figure 6** : Plot of the time evolution of the radius of gyration $R_g$ (top plot) and the aplanarity coefficient $a$ (bottom plot) for naked plasmids averaged over 8 simulations with $e_{SURF} = -0.08 e_{DNA}$ and different initial conditions. The smooth lines show the results of the fits with the expressions indicated in each vignette.

**Figure 7** : Plot of the characteristic times for adsorption, $\tau_a$ (disks), and equilibration, $\tau_R$ (lozenges), as a function of $-e_{SURF}/e_{DNA}$ for naked plasmids deposited on the charged surface.

**Figure 8 :** Plot of the time evolution of the radius of gyration $R_g$ (top plot) and the aplanarity coefficient $a$ (bottom plot) for the simulation with one pUC19 plasmid and 224 H-NS protein dimers and $e_{SURF} = -0.08 e_{DNA}$ also shown in the right column of Fig. 4. The two dot-dashed vertical lines labeled $t_{ad}$ and $t_{eq}$ indicate the times where the complex has completed adsorption (see the middle vignette in the right column of Fig. 4) and equilibration (see the bottom vignette in the right column of Fig. 4), respectively.

**Figure 9** : Plot of the time evolution of the radius of gyration $R_g$ (top plot) and the aplanarity coefficient $a$ (bottom plot) for complexes of plasmid and H-NS protein dimers averaged over 8 simulations with $e_{SURF} = -0.08\, e_{DNA}$ and different initial conditions. The smooth lines show the results of the fits with the expressions indicated in each vignette.

**Figure 10** : Plot of the characteristic times for adsorption, $\tau_a$ (upper plot), and equilibration, $\tau_R$ (lower plot), as a function of $-e_{SURF}/e_{DNA}$ for complexes of plasmid and H-NS protein dimers deposited on the charged surface.

**Figure 11 :** Probability distributions of the lifetime of H-NS bridges for $-e_{SURF}/e_{DNA} = 0.06$ (dashed line) and $-e_{SURF}/e_{DNA} = 0.10$ (solid line).

**Figure 12** : Plot of the time evolution of the number of *trans*-binding (top plot), *cis*-binding (middle plot), and dangling (bottom plot) H-NS protein dimers averaged over 8 simulations with $e_{SURF} = -0.08\, e_{DNA}$ and different initial conditions. The smooth lines show the results of the fits with the expressions indicated in each vignette.

**FIGURE 1**

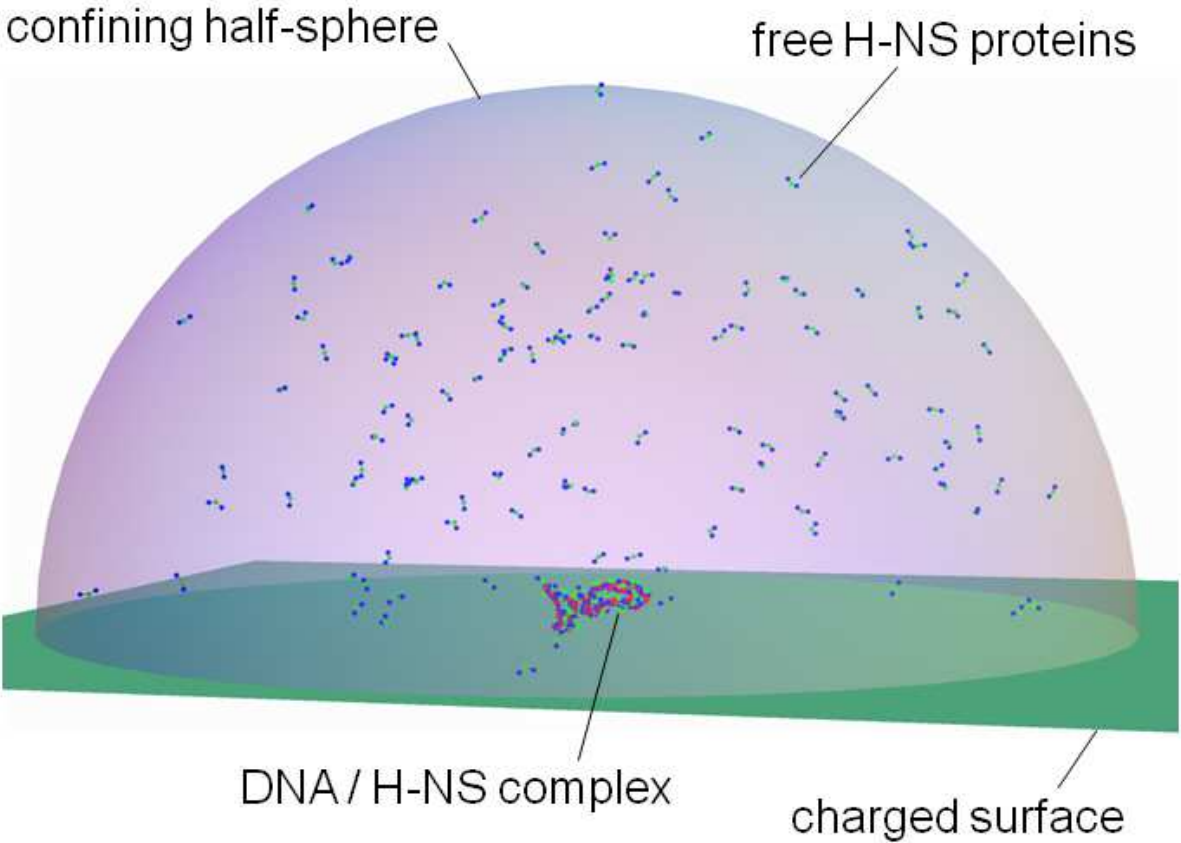

**FIGURE 2**

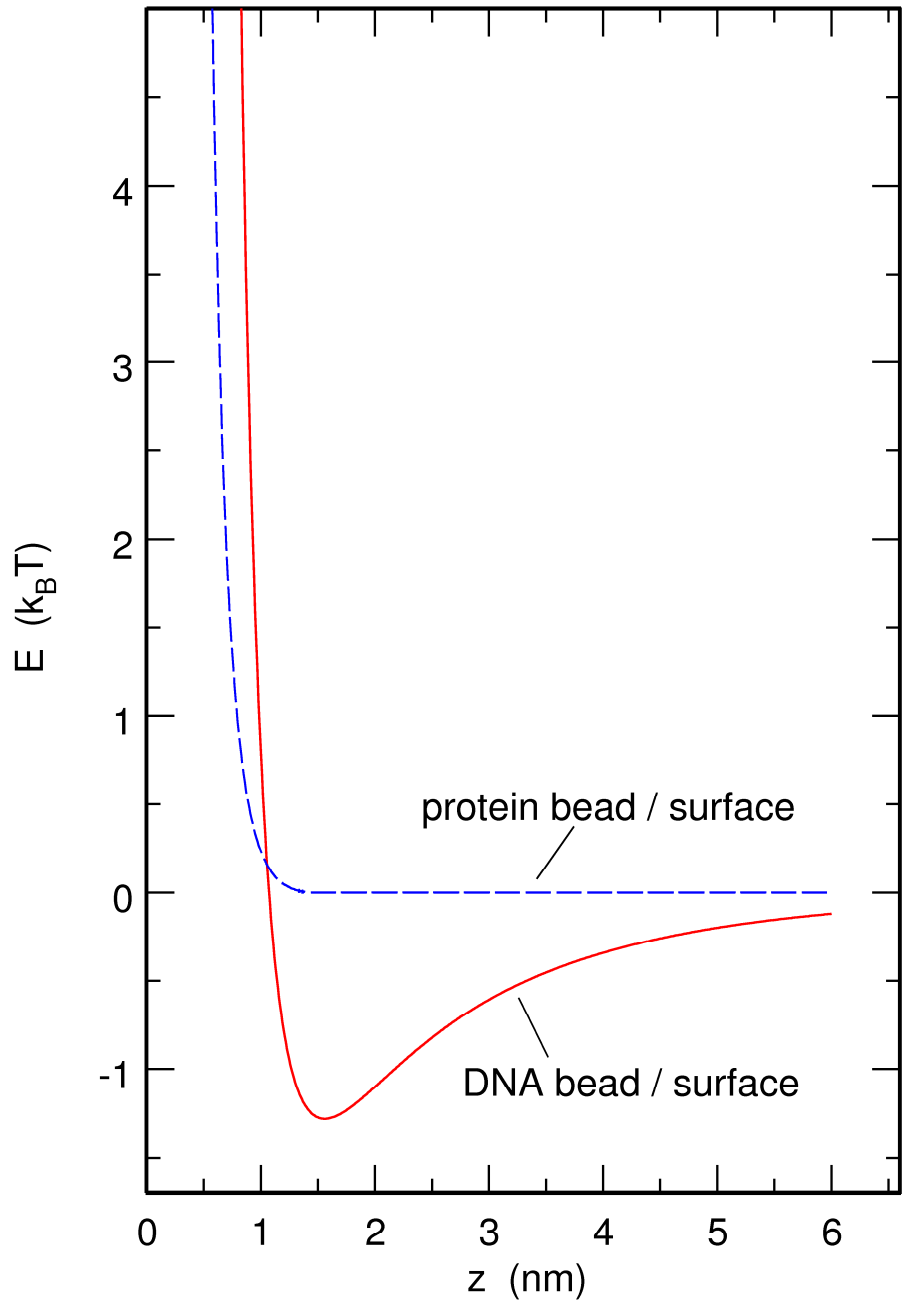

**FIGURE 3**

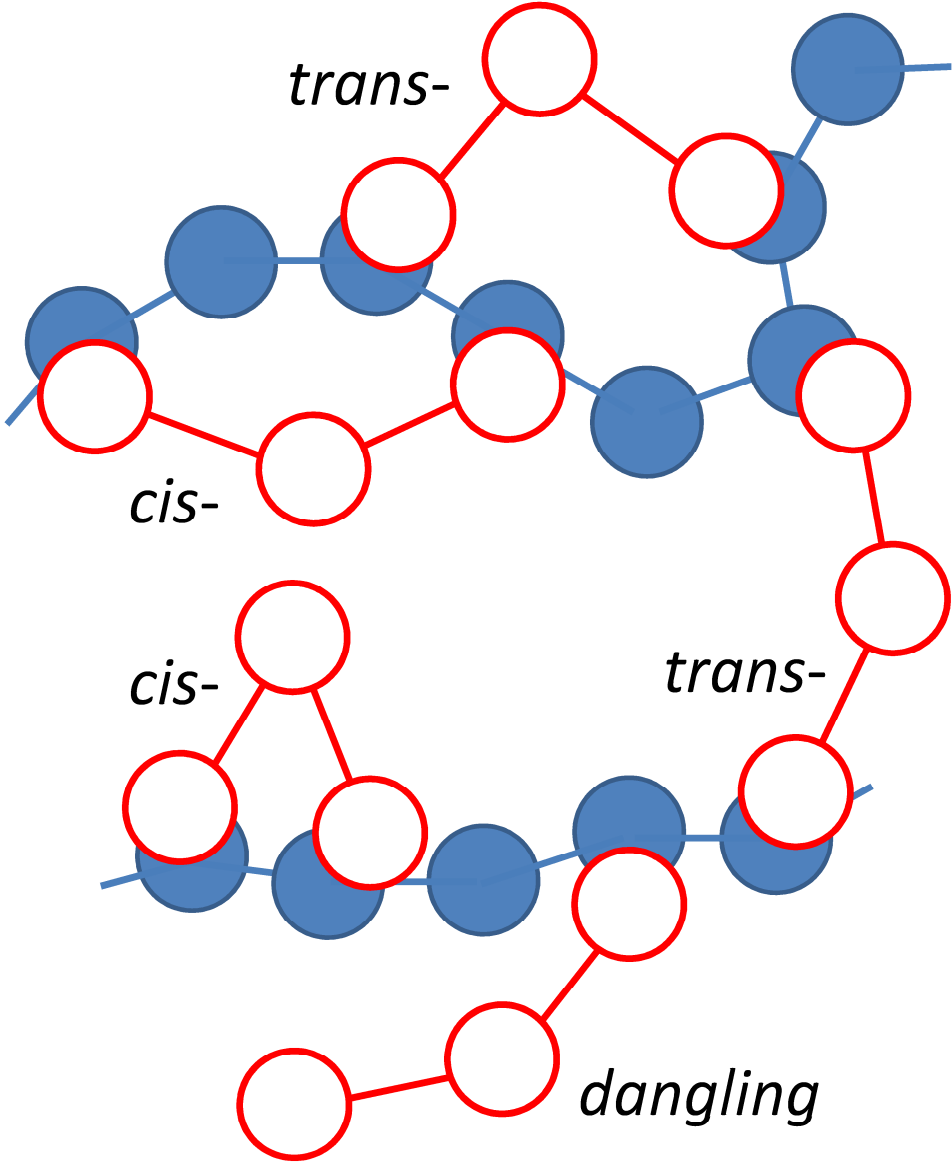

**FIGURE 4**

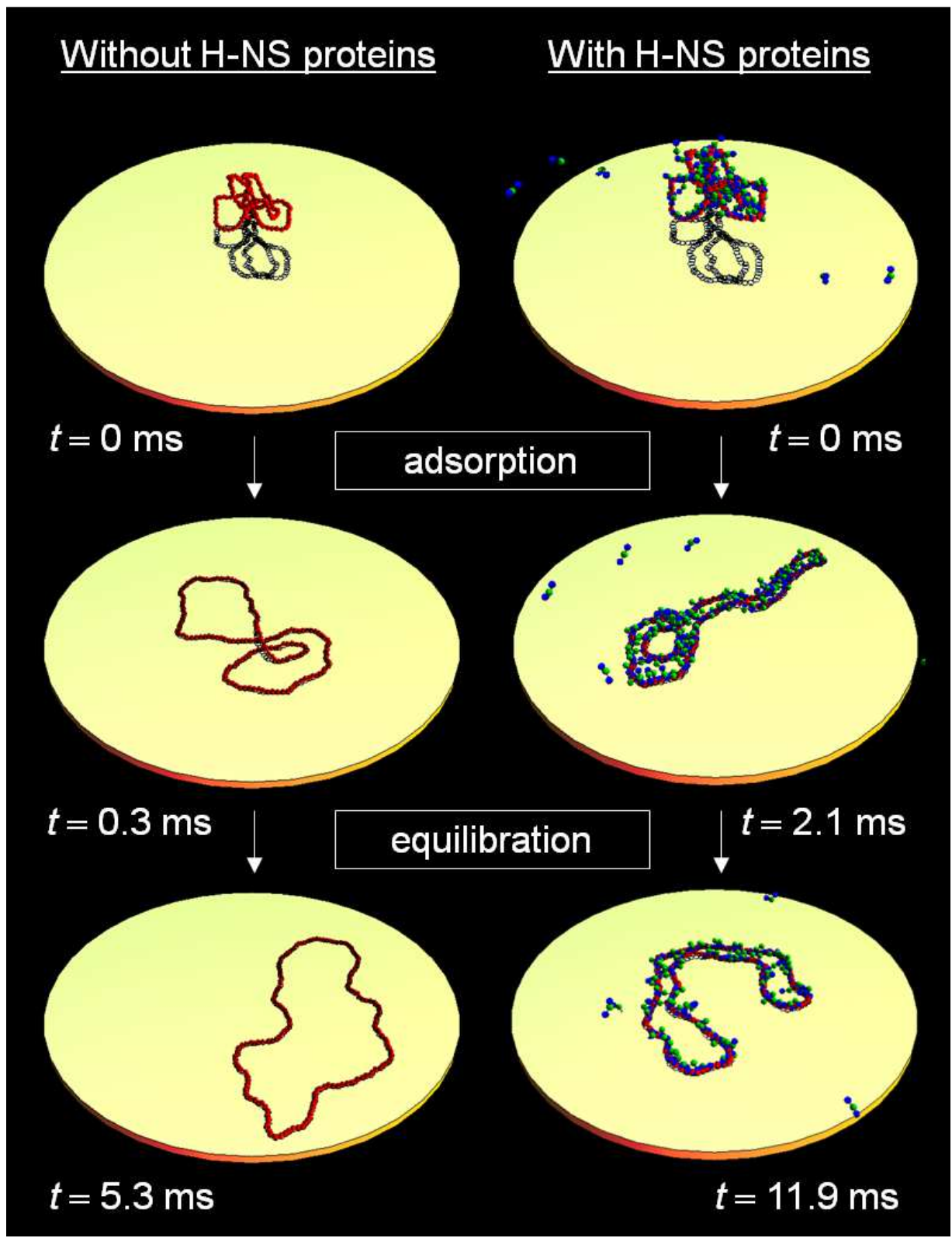

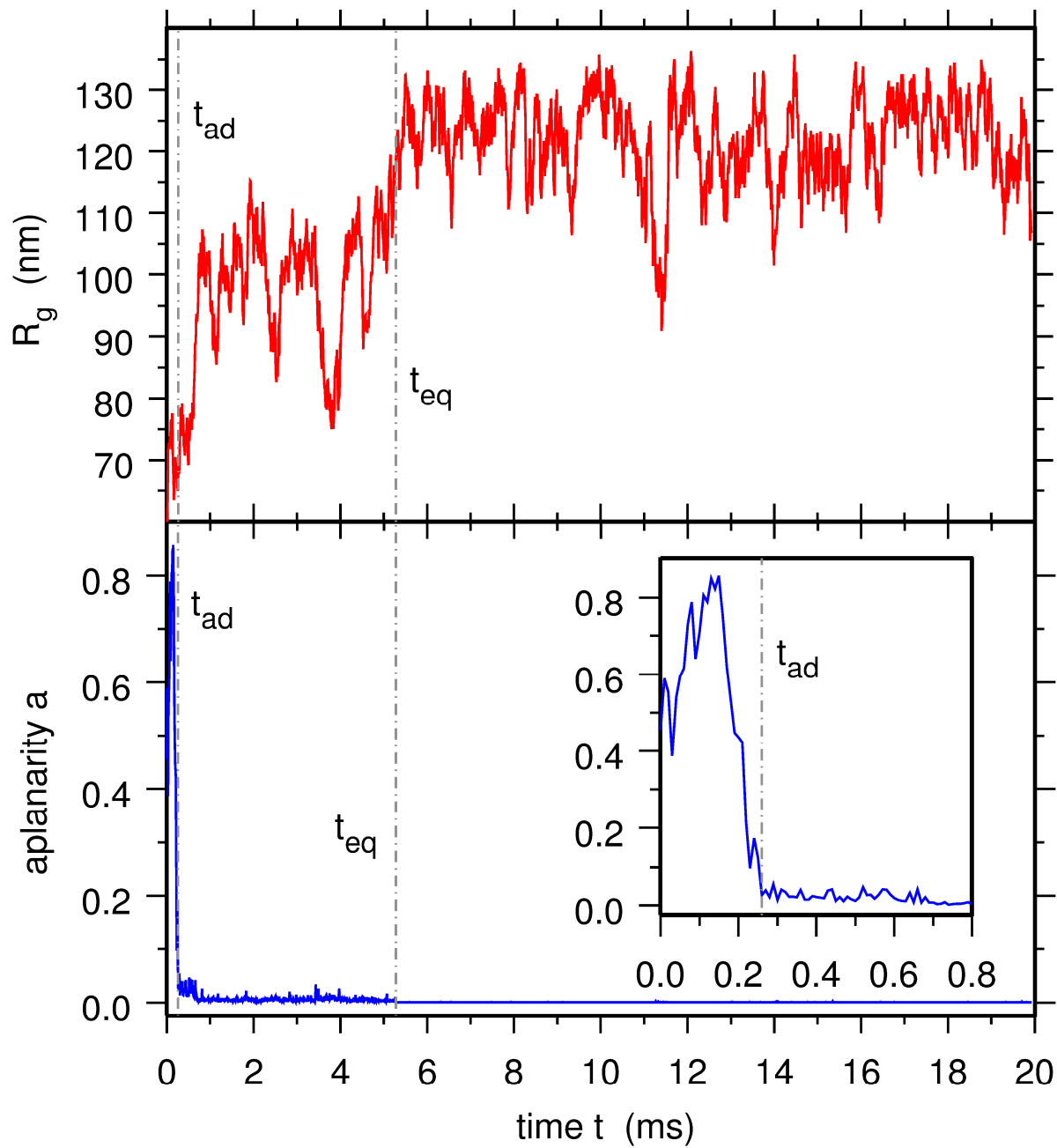

**FIGURE 6**

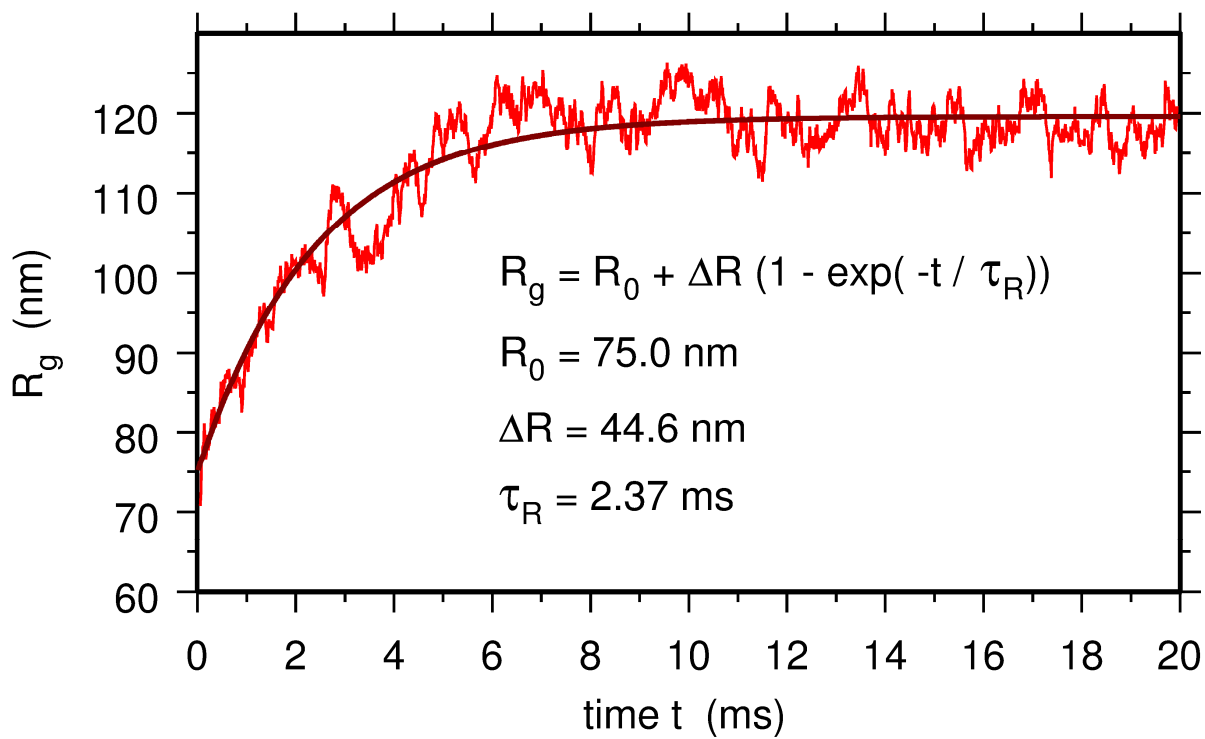

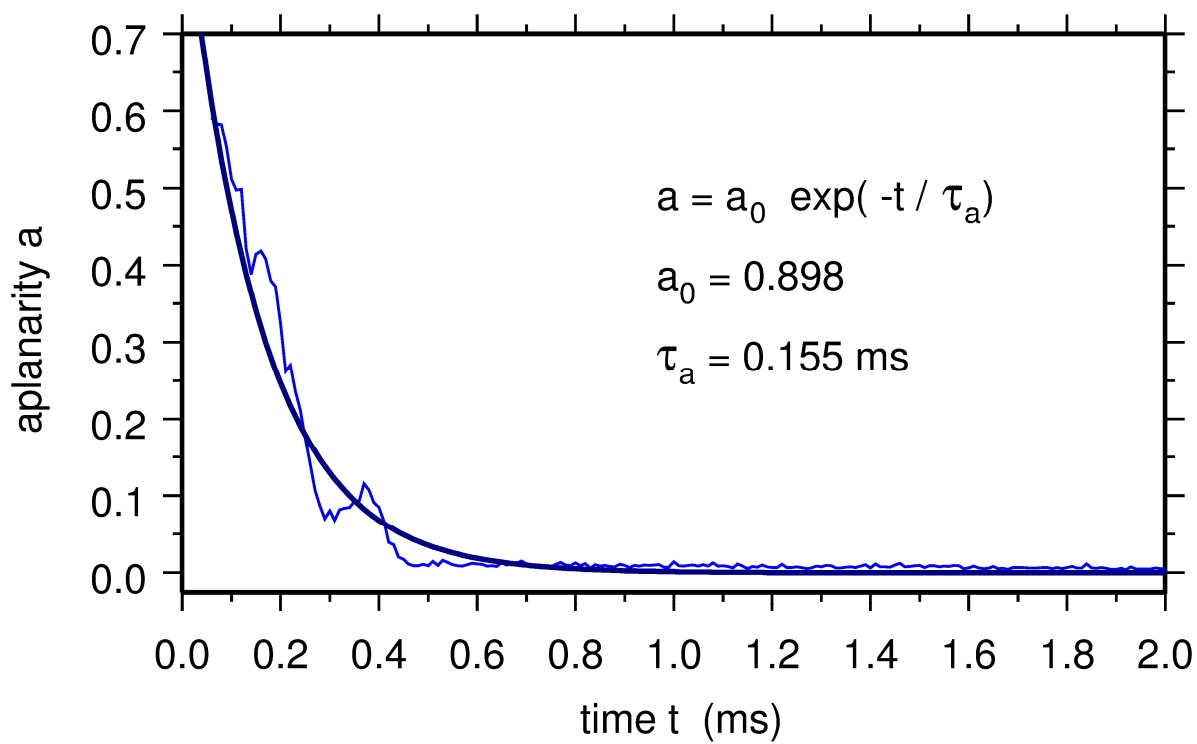

**FIGURE 7**

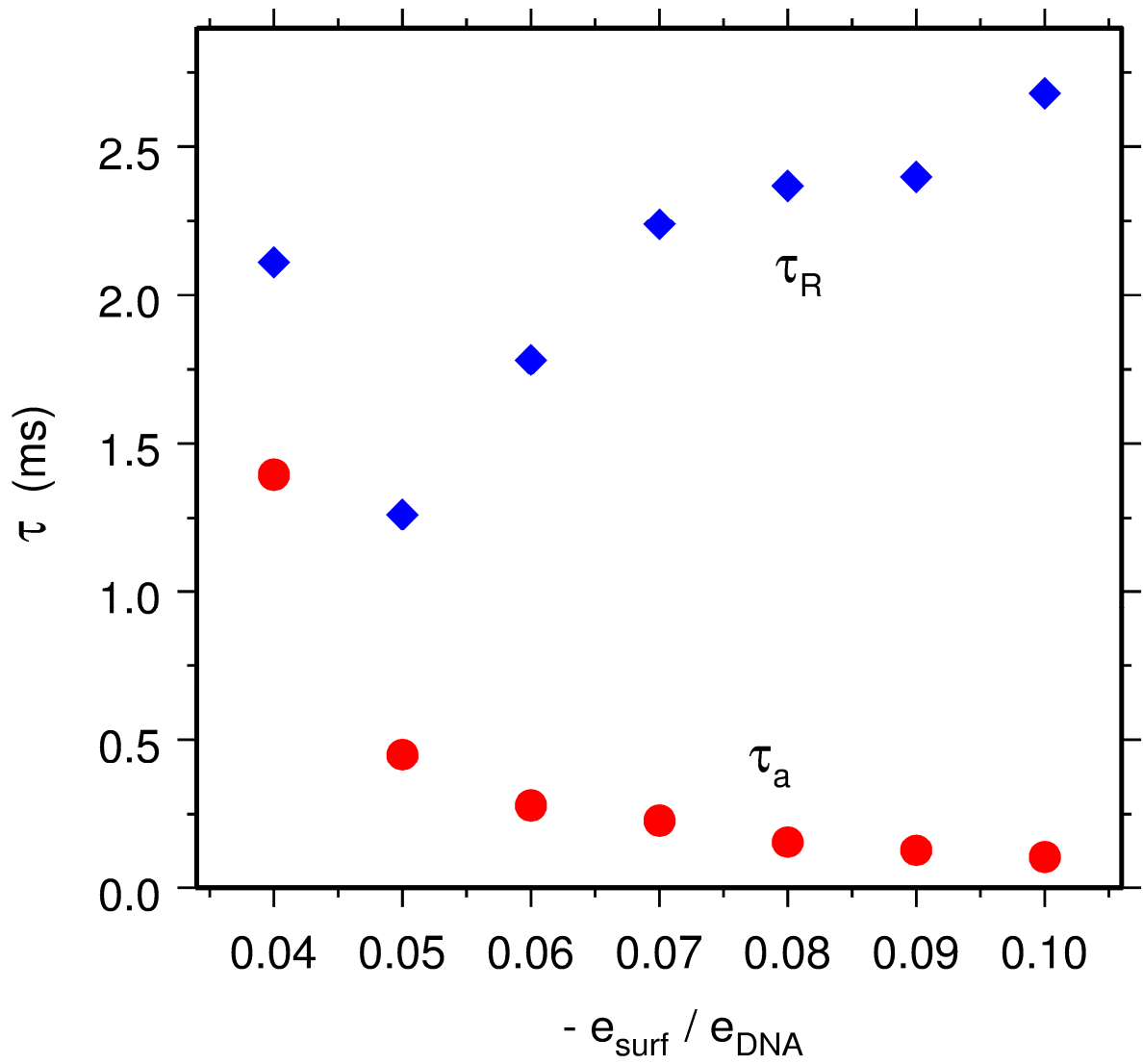

**FIGURE 8**

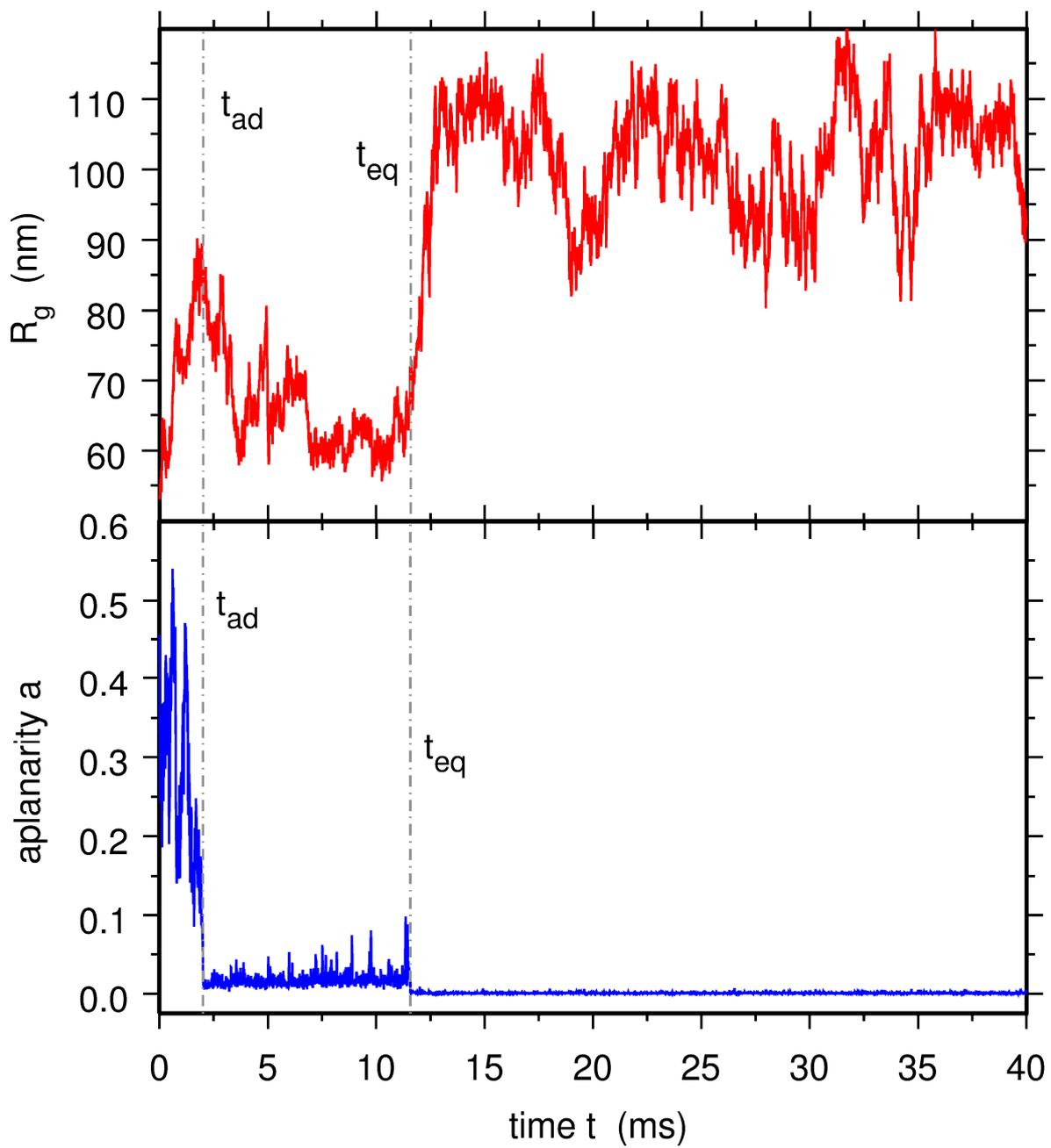

**FIGURE 9**

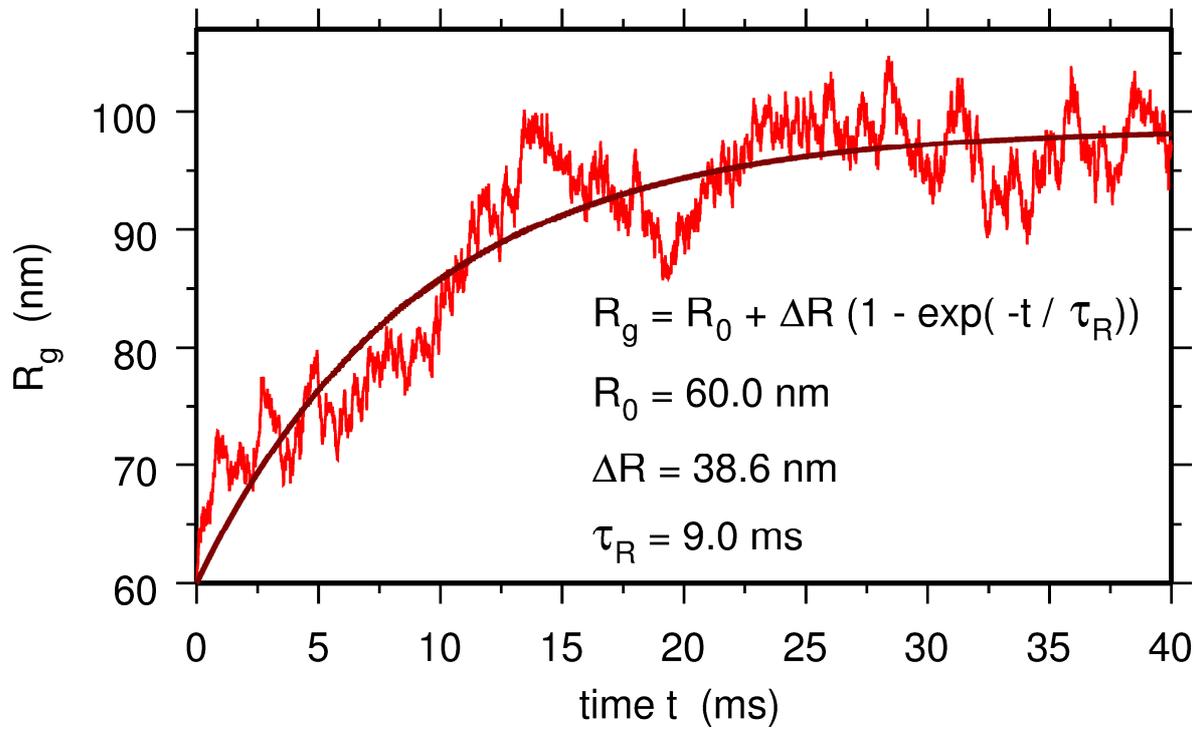

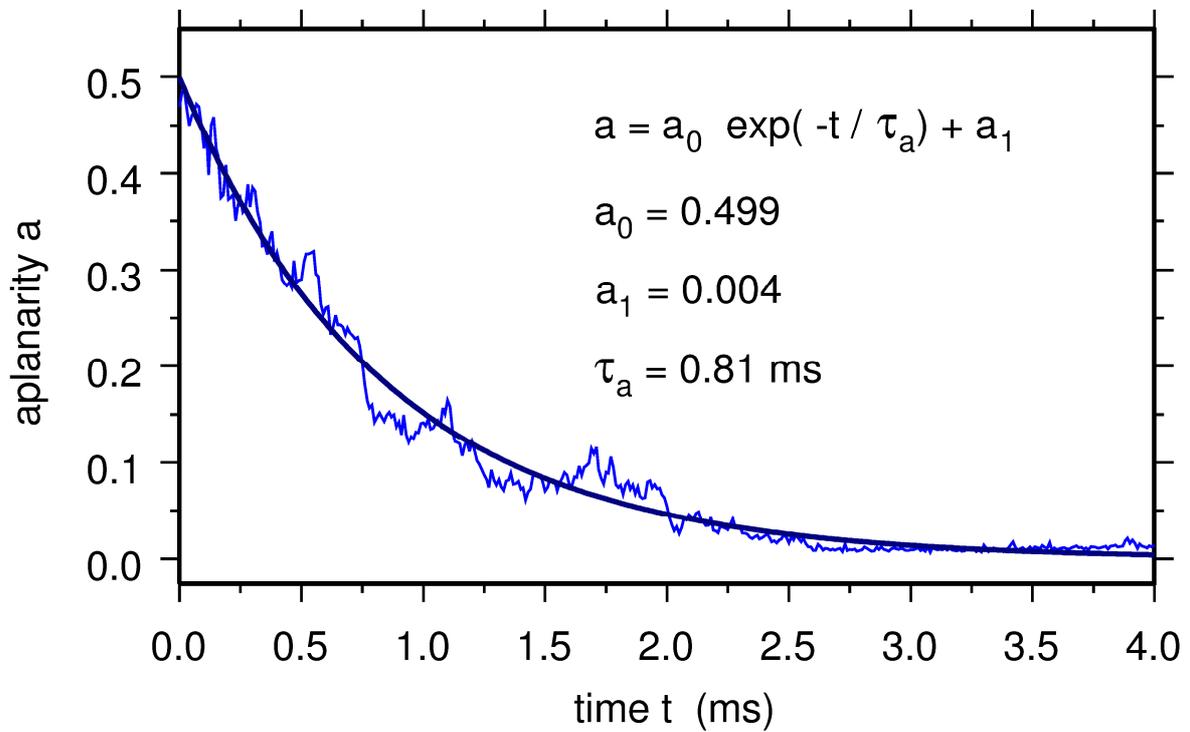

**FIGURE 10**

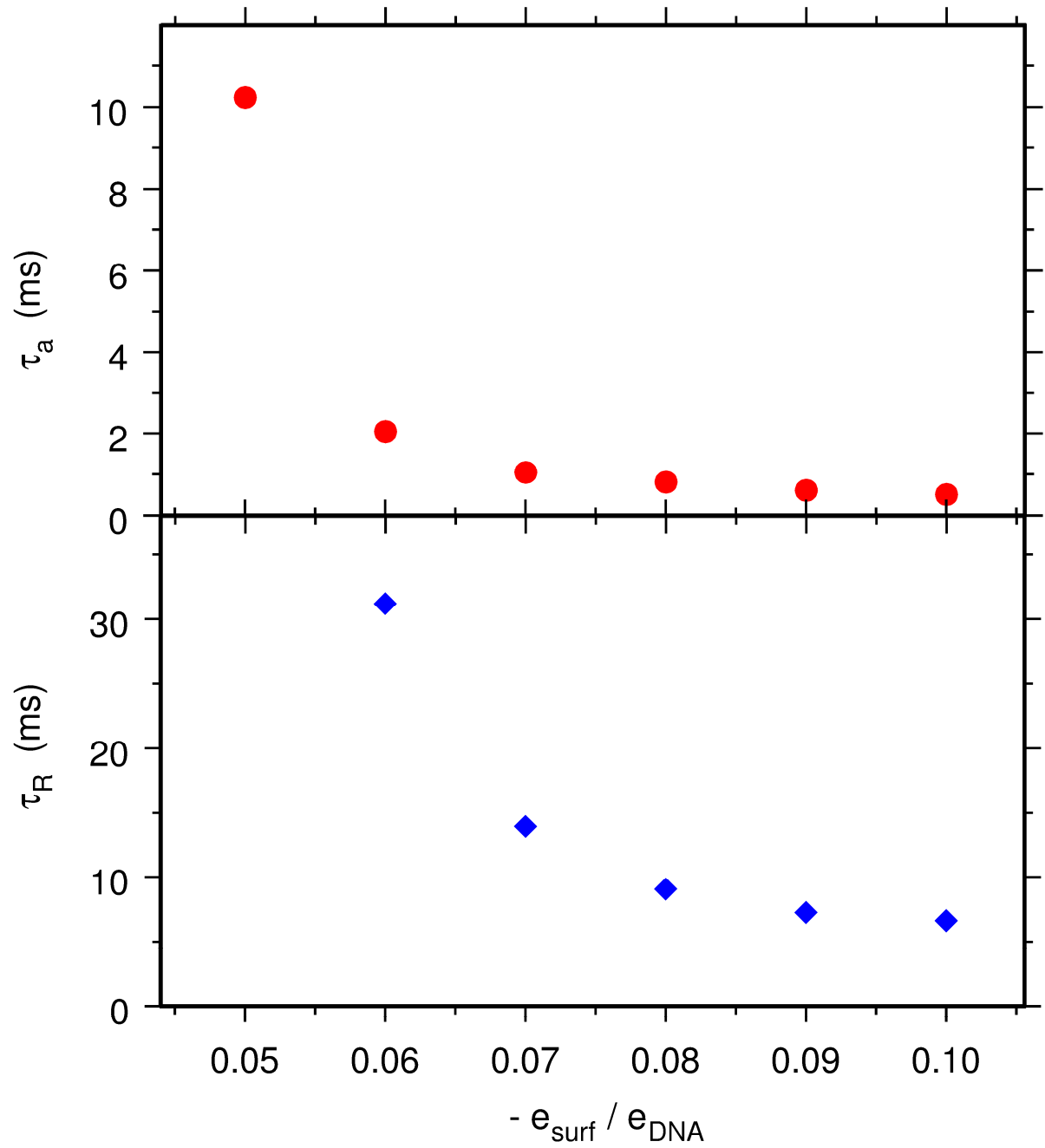

**FIGURE 11**

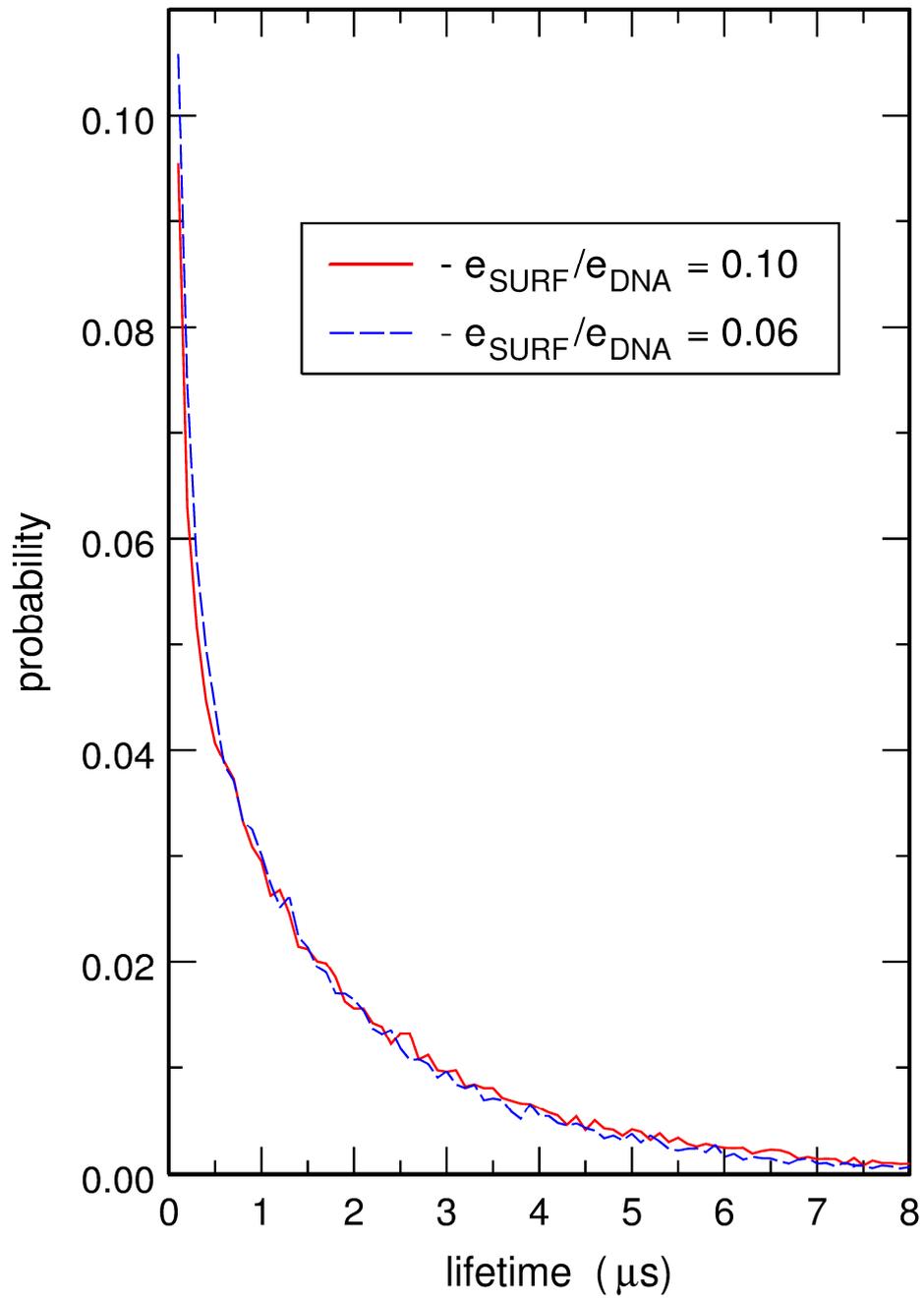

**FIGURE 12**

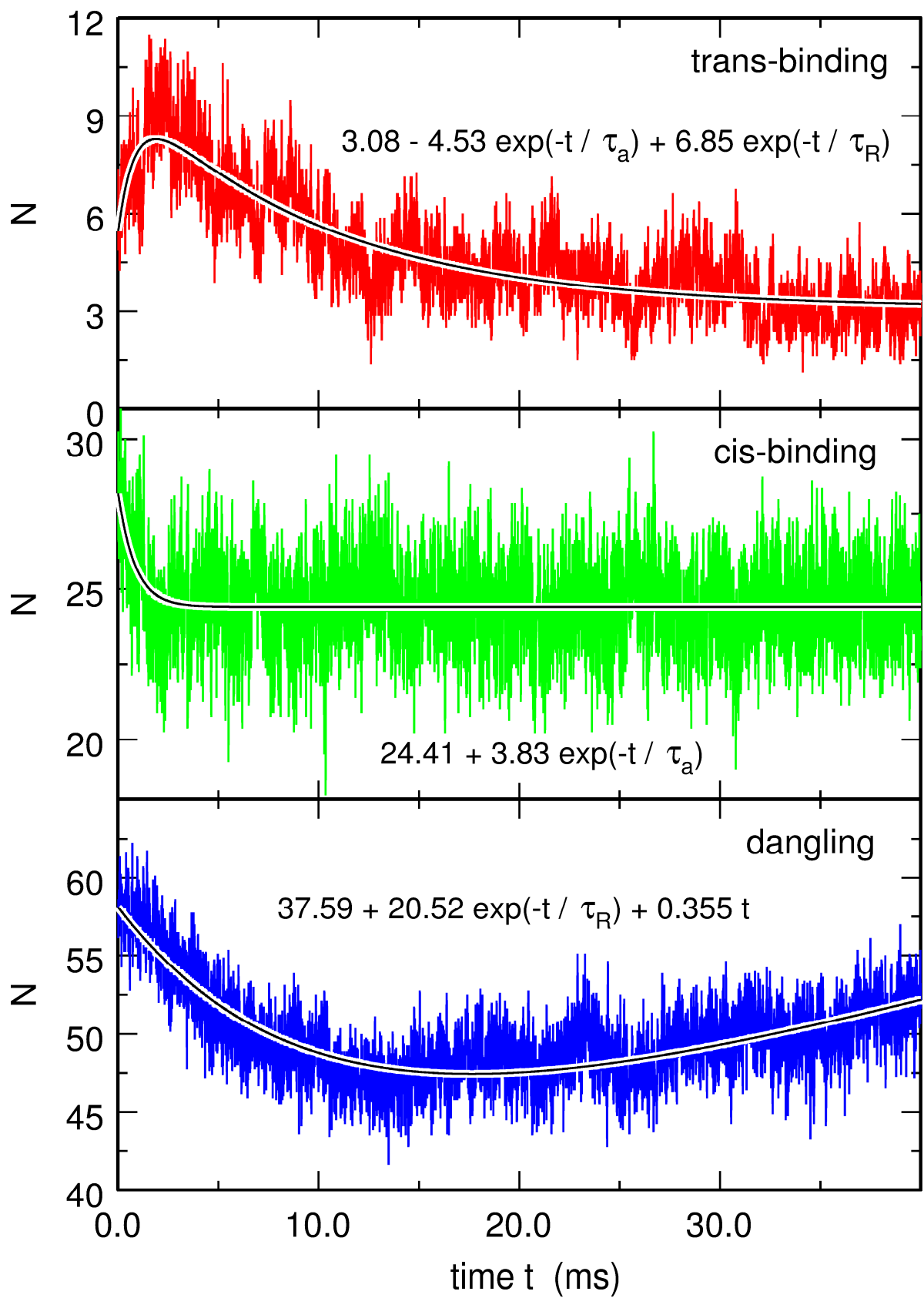

# Equilibration of complexes of DNA and H-NS proteins on charged surfaces: A coarse-grained model point of view.

## - Supplemental Material -

Marc JOYEUX
*Laboratoire Interdisciplinaire de Physique (CNRS UMR5588),*
*Université Joseph Fourier Grenoble 1, BP 87, 38402 St Martin d'Hères, France***EXPRESSION OF THE POTENTIAL ENERGY** $E_{\text{pot}}$

As shown in Fig. 1, the model consists of one pUC19 plasmid and $P=224$ H-NS dimers trapped between a charged surface and a confining half-sphere of radius $R_0 = 0.547$ μm, thus mimicking the experimental conditions of Dame and co-workers [1]. As in previous work [2-5], DNA is modelled as a chain of beads with hydrodynamic radius $a = 1.78$ nm separated at equilibrium by a distance $l_0 = 5.0$ nm. Each bead actually represents 15 base pairs, so that the pUC19 plasmid with 2686 bp is modelled as a cyclic chain of $n=179$ beads. It has been shown in Ref. [3] that this value of the hydrodynamic radius $a$ leads to correct diffusion coefficients for the DNA chains. Each bead has an effective charge $\approx -12\,\bar{e}$ placed at its centre, where $\bar{e}$ denotes the absolute charge of the electron. Each H-NS dimer is modelled as a chain of 3 beads with the same hydrodynamic radius $a = 1.78$ nm separated at equilibrium by a distance $L_0 = 7.0$ nm. An effective charge $-e_{\text{DNA}}/3 \approx 4\,\bar{e}$ is placed at the centre of each terminal bead and a charge $2e_{\text{DNA}}/3 \approx -8\,\bar{e}$ at the centre of each central bead

As discussed in the main text, the total potential energy of the system, $V_{\text{total}}$, is written in the form

$$V_{\text{total}} = E_{\text{pot}} + V_{\text{DNA/SURF}} + V_{\text{PROT/SURF}} \;, \tag{S0}$$

where $E_{\text{pot}}$ describes the potential energy of the system without the charged surface, $V_{\text{DNA/SURF}}$ the interactions between the plasmid molecule and the charged surface, and $V_{\text{PROT/SURF}}$ the interactions between H-NS molecules and the charged surface. The expressions of $V_{\text{DNA/SURF}}$ and $V_{\text{PROT/SURF}}$ are provided in Sec. II of the main text. The detailed expression of $E_{\text{pot}}$ was provided in the Supporting Material of Ref. [2] and is reported here for the sake of completeness.

$E_{\text{pot}}$ is taken as the sum of 5 terms

$$E_{\text{pot}} = V_{\text{DNA}} + V_{\text{PROT}} + V_{\text{PROT/PROT}} + V_{\text{DNA/PROT}} + V_{\text{wall}}, \quad (S1)$$

where $V_{\text{DNA}}$ and $V_{\text{PROT}}$ describe the potential energy of DNA and H-NS dimers, respectively, $V_{\text{PROT/PROT}}$ the interactions between H-NS dimers, $V_{\text{DNA/PROT}}$ the interactions between DNA and H-NS dimers, and $V_{\text{wall}}$ the repulsive wall that maintains H-NS dimers inside the sphere.

$V_{\text{DNA}}$ is in turn expressed, as in previous work [2-6], as a sum of three terms

$$\begin{aligned} V_{\text{DNA}} &= E_s + E_b + E_e \\ E_s &= \frac{h}{2} \sum_{k=1}^{n} (l_k - l_0)^2 \\ E_b &= \frac{g}{2} \sum_{k=1}^{n} \theta_k^2 \\ E_e &= e_{\text{DNA}}^2 \sum_{k=1}^{n-2} \sum_{K=k+2}^{n} H(\|\mathbf{r}_k - \mathbf{r}_K\|), \end{aligned} \quad (S2)$$

where $\mathbf{r}_k$ denotes the position of DNA bead $k$, $l_k = \|\mathbf{r}_k - \mathbf{r}_{k+1}\|$ the distance between two successive beads, $\theta_k = \arccos((\mathbf{r}_k - \mathbf{r}_{k+1})(\mathbf{r}_{k+1} - \mathbf{r}_{k+2})/(\|\mathbf{r}_k - \mathbf{r}_{k+1}\|\|\mathbf{r}_{k+1} - \mathbf{r}_{k+2}\|))$ the angle formed by three successive beads, and $H$ is the function defined according to

$$H(r) = \frac{1}{4\pi\varepsilon r} \exp\left(-\frac{r}{r_D}\right). \quad (S3)$$

$E_s$ is the bond stretching energy. This is a computational device without any biological meaning aimed at avoiding a rigid rod description. The stretching force constant is fixed at $h = 100 \, k_B T / l_0^2$, see the discussion in Ref. [3] for this choice for $h$. $E_b$ is the elastic bending potential. The bending rigidity constant, $g = 9.82 \, k_B T$, is chosen so as to provide the correct persistence length for DNA, which is 50 nm, equivalent to 10 beads [3,7]. $E_e$ is a Debye-Hückel potential, which describes repulsive electrostatic interactions between DNA beads [3,8,9]. In Eq. S3, $r_D = 3.07$ nm stands for the Debye length at 0.01 M salt concentration of monovalent ions [3] and $\varepsilon = 80 \, \varepsilon_0$ for the dielectric constant of the buffer. Electrostatic interactions between nearest-neighbours are not included in the expression of $E_e$ in Eq. S2, because these nearest-neighbour interactions are accounted for in the stretching and bending terms.

$V_{\text{PROT}}$ is similarly taken as the sum of stretching and bending contributions

$$V_{\text{PROT}} = E_s^{(P)} + E_b^{(P)}$$

$$E_s^{(P)} = \frac{h}{2} \sum_{j=1}^{P} (L_{j,1} - L_0)^2 + (L_{j,2} - L_0)^2 \qquad (S4)$$

$$E_b^{(P)} = \frac{G}{2} \sum_{j=1}^{P} \Theta_j^2,$$

where $L_{j,1}$, $L_{j,2}$, and $\Theta_j$, denote the distances between the terminal beads and the central bead and the angle formed by the three beads for the $j^{\text{th}}$ H-NS dimer. It can be estimated that $G$ is of the order of a few $k_B T$ for H-NS [2,10]. The value $G = 2\,k_B T$ was used for all simulations reported here.

The interaction between H-NS dimers, $V_{\text{PROT/PROT}}$, is taken as the sum of (attractive or repulsive) electrostatic terms and (repulsive) excluded volume terms, with the latter ones only contributing if the corresponding electrostatic interactions are attractive, i.e. between the terminal beads $m=1$ and $m=3$ of one dimer and the central bead $m=2$ of the other dimer

$$V_{\text{PROT/PROT}} = E_e^{(P/P)} + E_{ev}^{(P/P)}$$

$$E_e^{(P/P)} = \sum_{j=1}^{P} \sum_{m=1}^{3} \sum_{J=j+1}^{P} \sum_{M=1}^{3} e_{jm} e_{JM} H(\|\mathbf{R}_{jm} - \mathbf{R}_{JM}\|) \qquad (S5)$$

$$E_{ev}^{(P/P)} = \chi \sum_{j=1}^{P} \sum_{\substack{J=1 \\ J \neq j}}^{P} \left( \left|\frac{e_{j1} e_{J2}}{e_{\text{DNA}}^2}\right| F(\|\mathbf{R}_{j1} - \mathbf{R}_{J2}\|) + \left|\frac{e_{j3} e_{J2}}{e_{\text{DNA}}^2}\right| F(\|\mathbf{R}_{j3} - \mathbf{R}_{J2}\|) \right),$$

where $\mathbf{R}_{jm}$ denotes the position of bead $m$ of protein dimer $j$, $e_{jm}$ the charge placed at its centre, $\chi$ is a constant equal to $\chi = 0.15\,k_B T$, and $F$ is the function defined according to

$$\text{if } r \leq 2^{3/2} a : F(r) = 4\left(\left(\frac{2a}{r}\right)^4 - \left(\frac{2a}{r}\right)^2\right) + 1$$

$$\text{if } r > 2^{3/2} a : F(r) = 0. \qquad (S6)$$

It should be noted that electrostatic attractive interactions between the terminal beads of one H-NS dimer and the central bead of another H-NS dimer are too weak to allow for the formation of long-lived bonds between these two dimers.

The interaction between DNA and protein dimers, $V_{\text{DNA/PROT}}$, is similarly taken as the sum of (attractive or repulsive) electrostatic terms and (repulsive) excluded volume terms, with the latter ones only contributing if the corresponding electrostatic interactions are attractive, i.e. between DNA beads and terminal protein beads $m=1$ and $m=3$,

$$V_{\text{DNA/PROT}} = E_e^{(\text{DNA/P})} + E_{ev}^{(\text{DNA/P})}$$

$$E_e^{(\text{DNA/P})} = \sum_{j=1}^{P} \sum_{m=1}^{3} \sum_{k=1}^{n} e_{jm} e_{\text{DNA}} H(\|\mathbf{R}_{jm} - \mathbf{r}_k\|) \tag{S7}$$

$$E_{ev}^{(\text{DNA/P})} = \chi \sum_{j=1}^{P} \sum_{k=1}^{n} \left( \left|\frac{e_{j1}}{e_{\text{DNA}}}\right| F(\|\mathbf{R}_{j1} - \mathbf{r}_k\|) + \left|\frac{e_{j3}}{e_{\text{DNA}}}\right| F(\|\mathbf{R}_{j3} - \mathbf{r}_k\|) \right).$$

The confining sphere is large enough to fit the plasmid, whatever its current geometry and the centre of the sphere is furthermore adjusted at each time step to coincide with the centre of mass of the DNA molecule. DNA beads therefore cannot exit the sphere. Moreover, the repulsive wall $V_{\text{wall}}$ acts on the protein beads that move outside the radius of the sphere, $R_0$, and repels them back into the sphere. $V_{\text{wall}}$ is taken as a sum of repulsive terms

$$V_{\text{wall}} = 10 \, k_B T \sum_{j=1}^{P} \sum_{m=1}^{3} f(\|\mathbf{R}_{jm}\|) \,, \tag{S8}$$

where $f$ is the function defined according to

if $r \leq R_0$ : $f(r) = 0$

if $r > R_0$ : $f(r) = \left(\dfrac{r}{R_0}\right)^6 - 1$ . $\tag{S9}$

Within this model, interactions between DNA and H-NS dimers are essentially driven by the constant $\chi$ in Eq. S7. The value $\chi = 0.15 \, k_B T$ was chosen because it leads to a change in enthalpy $\Delta H$ of 11.1 $k_B T$ on forming a complex between DNA and an H-NS monomer, which is comparable to experimentally determined values [11].

# SUPPORTING REFERENCES

[1] Dame, R.T., C. Wyman, and N. Goosen. 2000. H-NS mediated compaction of DNA visualised by atomic force microscopy. *Nucleic Acids Res*. 28:3504-3510.

[2] Joyeux, M., and J. Vreede. 2013. A model of H-NS mediated compaction of bacterial DNA. *Biophys. J.* 104:1615-1622.

[3] Jian, H., A. Vologodskii, and T. Schlick. 1997. A combined wormlike-chain and bead model for dynamic simulations of long linear DNA. *J. Comp. Phys.* 136:168-179.

[4] Florescu, A.-M., and M. Joyeux. 2009. Description of non-specific DNA-protein interaction and facilitated diffusion with a dynamical model. *J. Chem. Phys.* 130:015103.

[5] Florescu, A.-M., and M. Joyeux. 2009. Dynamical model of DNA protein interaction: effect of protein charge distribution and mechanical properties. *J. Chem. Phys.* 131:105102.

[6] Florescu, A.-M., and M. Joyeux. 2010. Comparison of kinetic and dynamical models of DNA-protein interaction and facilitated diffusion. *J. Phys. Chem. A* 114:9662-9672.

[7] Frank-Kamenetskii, M.D., A.V. Lukashin, and V.V. Anshelevich. 1985. Torsional and bending rigidity of the double helix from data on small DNA rings. *J. Biomol. Struct. Dynam.* 2:1005-1012.

[8] Stigter, D. 1977. Interactions of highly charged colloidal cylinders with applications to double-stranded DNA. *Biopolymers* 16:1435-1448.

[9] Vologodskii, A.V., and N.R. Cozzarelli. 1995. Modeling of long-range electrostatic interactions in DNA. *Biopolymers* 35:289-296.

[10] Sivaramakrishnan S., J. Sung, M. Ali, S. Doniach, H. Flyvbjerg, and J.A. Spudich. 2009. Combining single-molecule optical trapping and small-angle X-ray scattering measurements to compute the persistence length of a protein ER/K α-helix. *Biophys. J.* 97:2993-2999.

[11] Ono, S., M.D. Goldberg, T. Olsson, D. Esposito, J.C.D. Hinton, and J.E. Ladbury. 2005. H-NS is a part of a thermally controlled mechanism for bacterial gene regulation. *Biochem. J.* 391:203-213.